\documentclass[preprint,eqsecnum,preprintnumbers,nofootinbib,byrevtex,prd,aps,showpacs,showkeys,groupedaddress,floatfix]{revtex4}
\usepackage{bm}
\usepackage[dvips]{graphicx}
\usepackage{graphics}
\usepackage{graphicx}
\usepackage{epsfig}
\usepackage{amssymb}
\usepackage{amsmath}
\begin{document}

\title{SINGLE MESON PRODUCTION IN  PHOTON-PHOTON COLLISIONS AND INFRARED RENORMALONS}
\author{A.~I.~Ahmadov$^{1,2}$~\footnote{ahmadovazar@yahoo.com}}
\author{Coskun ~Aydin$^{3}$~\footnote{coskun@ktu.edu.tr}}
\author{E.~A.~Dadashov$^{4}$}
\author{Sh.~M.~Nagiyev$^{4}$}%
\affiliation{$^{1}$
Institut f\"{u}r Theoretische Physik E\\
RWTH Aachen University, D-52056 Aachen, Germany}
\affiliation{$^{2}$ Department of Theoretical Physics, Baku State
University  \\ Z. Khalilov Street 23, AZ-1148, Baku, Azerbaijan}
\affiliation{$^{3}$ Department of Physics, Karadeniz Technical
University, 61080, Trabzon, Turkey } \affiliation{$^{4}$
Institute of Physics of Azerbaijan National Academy of Sciences\\
 H. Javid Avenue, 33, AZ-1143, Baku, Azerbaijan}
\begin{abstract}
In this article, we investigate the contribution of the higher-twist
Feynman diagrams to the large-$p_T$  inclusive single meson
production cross section in photon-photon collisions and present the
general formulas for the higher-twist differential cross sections in
case of the running coupling and frozen coupling approaches. The
structure of infrared renormalon singularities of the higher-twist
subprocess cross section and the resummed expression (the Borel sum)
for it are found. We compared the resummed higher-twist cross
sections with the ones obtained in the framework of the frozen
coupling approach and leading-twist cross section. We obtain, that
ratio
$R=(\Sigma_{M^{+}}^{HT})^{res}$/$(\Sigma_{M^{+}}^{HT})^{0}$, for
all values of the transverse momentum $p_{T}$ of the meson
identically equivalent to ratio
$r$=($\Delta_{M}^{HT})^{res}$/$(\Delta_{M}^{HT})^{0}$. It is
shown that the resummed result depends on the choice of the meson
wave functions used in calculation. Phenomenological effects of the
obtained results are discussed.
\end{abstract}

\maketitle

\section{\bf Introduction}

Exclusive processes involving large momentum transfer are among the
most interesting and challenging test of quantum chromodynamics
(QCD).
The framework for analyzing such processes within the context
of perturbative  QCD (pQCD) has been developed by Brodsky and Lepage
[1,2], Efremov and Radyshkin [3], and Duncan and Mueller [4]. They
have demonstrated, to all orders in perturbation theory, that
exclusive amplitudes involving large momentum transfer factorize
into a convolution of a process-independent and perturbatively
incalculable distribution amplitude, one for each hadron involved in
the amplitude, with a process-dependent and perturbatively
calculable hard-scattering amplitude.

The hadronic wave function in terms of quark and gluon degrees of
freedoms plays an important role in QCD process predictions. For
example, knowledge of the wave function allows to calculate
distribution amplitudes and structure functions or conversely these
processes can give phenomenological restrictions on the wave
functions.

During the last few years, a great deal of progress has been made in
the investigation of the properties of hadronic wave
functions[5-17].

The standard approach to distribution amplitudes, which is due to
Brodsky and Lepage[14], considers the hadron's parton decomposition
in the infinite momentum frame. A conceptually different, but
mathematically equivalent formalism  is the light-cone
quantization[15]. Either way, power-suppressed contributions to
exclusive processes in QCD, which are commonly referred to as
higher-twist corrections. The higher-twist approximation  describes
the multiple scattering of a parton as power corrections to the
leading-twist cross section.

Among the fundamental predictions of QCD are asymptotic scaling laws
for large-angle exclusive processes [18-22]. QCD counting rules were
formalized in Refs.[19,20].

The frozen coupling constant approach can be applied for
investigation, not only exclusive processes, but also for the
calculation of higher-twist contributions to some inclusive
processes, for example as large -$p_{T}$ meson photoproduction [23],
two-jet+meson production in the electron-positron annihilation [24].
In the works [24,25] for calculation of integrals, such as

\begin{equation}
I\sim \int\frac{\alpha_{s}(\hat {Q}^2)\Phi(x,\hat{Q}^2)}{1-x}dx
\end{equation}

the frozen coupling constant approach was used. According to
Ref.[25] should be noted that in pQCD calculations the argument of
the running coupling constant in both,  the renormalization and
factorization scale $\hat{Q}^2$ should be taken equal to the
square of the momentum transfer of a hard gluon in a corresponding
Feynman diagram. But defined in this way, $\alpha_{s}(\hat{Q}^2)$
suffers from infrared singularities. For example in our work [26],
$\hat{Q}^2$ equals to $x_{2}\hat{s}$ and $-x_{1}\hat{u}$, where
$\hat{s}$, $\hat{u}$ are the subprocess's Mandelstam invariants.
Therefore, in the soft regions $x\rightarrow 0$, integrals (1.1)
diverge and for their calculation some regularization methods of
$\alpha_{s}(Q^2)$ in these regions are needed. In Ref.[27], the
authors investigated the phenomenology of infrared renormalons in
inclusive processes. The dispersive approach has been devised to
extend properly modified perturbation theory calculations towards
the low-energy region [28]. Connections between power corrections
for the three Deep Inelastic Scattering sum rules have also been
explored in [29].

Investigation of the infrared renormalon effects in various
inclusive and exclusive processes is one of the most important and
interesting problems in the perturbative QCD. As we know the  word
"renormalon" first appeared in Ref.[30]. A singularity in the Borel
parameter- is called a renormalon. It is known that infrared
renormalons are responsible for factorial growth of coefficients in
perturbative series for the physical quantities. But, these
divergent  series can be resummed by means of the Borel
transformation [30] and the principal  value prescription [31], and
effects of infrared renormalons can be taken into account by a
scale-setting procedure
$\alpha_{s}(Q^2)\rightarrow\alpha_{s}(exp(f(Q^2))Q^2)$ at the
one-loop order results. Technically, all-order resummation of
infrared renormalons corresponds to the calculation of the one-loop
Feynman diagrams with the running coupling constant
$\alpha_{s}(-k^2)$ at the vertices or, alternatively, to calculation
of the same diagrams with nonzero gluon mass. Studies of infrared
renormalon problems have also opened new prospects for evaluation of
power-suppressed corrections to processes characteristics [32].
Power corrections can also be obtained by means of the Landau-pole
free expression for the QCD coupling constant. The most simple and
elaborated variant of the dispersive approach, the Shirkov and
Solovtsov analytic perturbation theory, was formulated in Ref.[33].
The $k_T$ factorization theorem has been widely applied to inclusive
and exclusive processes in perturbative QCD. This theorem holds for
simple processes, such as deeply inelastic scattering (DIS) and
Drell-Yan production[34].

A full twist 3 treatment of $\rho$-electroproduction in $k_T$
-factorisation is possible[35]. It relies on the computation of the
$\gamma_{T}^\ast-\rho_T $ impact factor at twist 3 including
consistently all twist 3 contributions, i.e. 2-parton and 3-parton
correlators. This gives a gauge invariant impact factor, and an
amplitude which is free of end-point singularities due to the
presence of $k_T$ .

 An additional, general property is that the singularities in $B[u]$
 occur at integer-and sometimes half integer-values of $u$. This
 corresponds to the fact that alternative definitions of the sum of
 the series differ by integer-or half integer-power of
 $\Lambda^{2}/Q^2$. These ambiguities must be cancelled by
 nonperturbative power corrections, and they can therefore serve as
 a perturbative probe of such effects. In the absence of an operator
 product expansion, the renormalon technique often provides a unique
 window into the nonperturbative regime: by identifying the
 ambiguities in summing the perturbative series one learns about the
 parametric dependence of power corrections on the hard scales and
 about their potential size[36].

By taking these points into account, it may be argued that the
analysis of the higher-twist effects  on the dependence of the meson
wave function  in single pseudoscalar and vector meson production at
photon-photon collisions by the running coupling (RC) approach are
significant from both theoretical and experimental points of view.

In this work we will apply the running coupling approach[37] in
order to compute effects of the infrared renormalons on the meson
production in photon-photon collisions. This approach was employed
also in our work[38] for calculation of the single meson
production in proton-proton collisions.

Photon-photon collisions represent a very useful tool for the
study of hadron production. Basically, the more attractive feature
is the simple, clean initial state, involving only QED
interactions, which allows one to concentrates on the final
hadronic state. This way, in fact, some of the more clean tests
for pQCD models were proposed [14]. It is well known that
exclusive $\gamma\gamma \to hadron$ processes can be studied in
the $e^{+}e^{-}$ colliders, particularly $\gamma^{*}\gamma^{*}$
processes, play a spesial role in QCD [39], since their analysis
is under much better control than the calculation of hadronic
processes, which requre the input of non-perturbative hadronic
structure functions or wave functions.

A precise measurement of the inclusive charged meson production
cross section at $\sqrt s=183 GeV$ and $\sqrt s=209 GeV$ is
important for the photon-photon collisions program at the
International Linear Collider (ILC). The results of our calculations
are based on the photon-photon collisions at $\sqrt s=183 GeV$ and
$\sqrt s=209 GeV$.

The higher-twist contributions to high-$p_T$ inclusive meson
production in two-photon collisions, a single meson inclusive
photoproduction and jet photoproduction cross sections were
studied by various authors [40-42]. As experiments examining
high-$p_T$ particle production in two-photon collisions are
improved, it becomes important to reassess the various
contributions which arise in quantum chromodynamics. Predicting
for the higher-twist contributions, originally obtained in Ref.43,
may now be refined using the exclusive-process QCD formalism
devoloped in [44]. Another important aspect of this study is the
choice of the meson model wave functions. In this respect, the
contribution of the higher-twist Feynman diagrams to a single
meson production cross section in photon-photon collisions has
been computed by using various meson wave functions. Also, the
leading and higher-twist contributions have been estimated and
compared to each other. Within this context, this paper is
organized as follows: in Sec. \ref{ht}, we provide some formulas
for the calculation of the contribution of the high twist
diagrams. In Sec. \ref{ir} we present formulas and an analysis of
the higher-twist effects on the dependence of the meson wave
function by the running coupling constant approach. In Sec.
\ref{lt}, we provide the formulas for the calculation of the
contribution of the leading-twist diagrams and in Sec.
\ref{results}, we present the numerical results for the cross
section and discuss the dependence of the cross section on the
meson wave functions. We state our conclusions in section
\ref{conc}.

\section{CONTRIBUTION OF THE HIGH TWIST DIAGRAMS}\label{ht}
The higher-twist Feynman diagrams, which describe the subprocess
$\gamma q \to M q $ contributes to $\gamma\gamma \to MX$ for the
meson production in the photon-photon collision are shown in
Fig.1(a). The amplitude for this subprocess can be found by means
of the Brodsky-Lepage formula [45]

\begin{equation}
M(\hat s,\hat
t)=\int_{0}^{1}{dx_1}\int_{0}^{1}dx_2\delta(1-x_1-x_2)\Phi_{M}(x_1,x_2,Q^2)T_{H}(\hat
s,\hat t;x_1,x_2).
\end{equation}

In Eq.(2.1), $T_H$ is  the sum of the graphs contributing to the
hard-scattering part of the subprocess. The hard-scattering part for
the subprocess under consideration is $\gamma q_1 \to
(q_{1}\overline{q}_{2})q_2$, where a quark and antiquark form a
pseudoscalar, color-singlet state $(q_1\bar{q}_2)$. Here
$\Phi(x_1,x_2,Q^2)$ is the meson wave function, i.e., the probability
amplitude for finding the valence $q_1\bar{q}_2$ Fock state in the
meson carry fractions $x_1$ and $x_2$, $x_1+x_2=1$. Remarkably, this
factorization is gauge invariant and only requires that the momentum
transfers in $T_H$ be large compared to the intrinsic mass scales of
QCD. Since the distribution amplitude and the hard scattering
amplitude are defined without reference to the perturbation theory,
the factorization is valid to leading order in $1/Q$, independent of
the convergence of perturbative expansions.
The Hard-scattering amplitude $T_H$ can be calculated in perturbation
theory and represented as a series in the QCD running coupling
constant $\alpha_s(Q^2)$.

The $q_{1}\overline{q}_{2}$ spin state used in computing
$T_H$ may be written in the form

\begin{equation}
\sum_{s_{1},s_{2}}
\frac{u_{s_1}({x}_{1}p_{M})\overline{v}_{s_{2}}({x}_{2}p_{M})}{\sqrt{x_1}
\sqrt{x_2}}\cdot N_{s_{1}s_{2}}^s=\left\{\begin{array}{ccc}
\frac{{\gamma}_{5}\hat {p}_{\pi}}{\sqrt{2}},\,\,\pi,\\\frac{\hat
{p}_{M}}{\sqrt{2}},\,\,\rho_L\,\,helicity \, 0,\\
\mp\frac{{\varepsilon}_{\mp}\hat {p}_{M}}{\sqrt{2}},\,\,
\rho{_T}\,\,helicity \pm1,\end{array}\right.
\end{equation}

where $\varepsilon_{\pm}=\mp(1/\sqrt{2})(0,1,\pm i,0)$ in a frame
with $(p_M)_{1,2}=0$ and the $N_{s_{1}s_{2}}^s$ project out a state
of spins $s$, and $p_{M}$ is the four-momentum of the final meson.
In our calculation, we have neglected the meson mass. Turning to
extracting the contributions of the higher-twist subprocesses, there
are many kinds of leading-twist subprocesses in $\gamma\gamma$
collisions as the background of the higher-twist subprocess $\gamma
q \to Mq$, such as $\gamma+\gamma \to q+\overline{q}$. The
contributions from these leading-twist subprocesses strongly depend
on some phenomenological factors, for example, quark and gluon
distribution functions in meson and fragmentation functions of
various constituents \emph{etc}. Most of these factors have not been
well determined, neither theoretically nor experimentally. Thus they
cause very large uncertainty in the computation of the cross section
of process $\gamma\gamma \to MX$. In general, the magnitude of this
uncertainty is much larger than the sum of all the higher-twist
contributions, so it is very difficult to extract the higher-twist
contributions.

The Mandelstam invariant variables for subprocesses
$\gamma q\to Mq$ are defined as

\begin{equation}
\hat s=(p_1+p_{\gamma})^2,\quad \hat t=(p_{\gamma}-p_{M})^2,\quad \hat
u=(p_1-p_{M})^2.
\end{equation}

We have aimed to calculate the meson production cross section and
to fix the differences due to the use of various meson model
functions. We have used seven different wave functions: the
asymptotic wave function (ASY), the Chernyak-Zhitnitsky wave
function [6,9], the wave function in which two nontrivial
Gegenbauer coefficients $a_2$ and $a_4$ have been extracted from
the CLEO data on the $\gamma\gamma^{\star} \to \pi^0$ transition
form factor [46], the Braun-Filyanov pion wave functions [11] and
the Bakulev-Mikhailov-Stefanis pion wave function[47]. It should
be noted that the wave functions of pions also are developed in
Refs.[48-50] by the Dubna group. For $\rho$- meson wave function
we used the Ball-Braun wave function[51].

$$
\Phi_{asy}(x)=\sqrt{3}f_{\pi}x(1-x),\quad
\Phi_{L(T)}^{asy}(x)=\sqrt{6}f_{\rho}^{L(T)}x(1-x),\\
$$
$$
\Phi_{CZ}(x,\mu_{0}^2)=\Phi_{asy}(x)\left[C_{0}^{3/2}(2x-1)+\frac{2}{3}C_{2}^{3/2}(2x-1)\right],
$$
$$
\Phi_{L(T)}^{\rho}(x,\mu_{0}^2)=\Phi_{L(T)}^{asy}(x)\left[C_{0}^{3/2}(2x-1)+0.18(0.2)\frac{2}{3}C_{2}^{
3/2}(2x-1)\right],
$$
$$
\Phi_{BMS}(x,\mu_{0}^2)=\Phi_{asy}(x)\left[C_{0}^{3/2}(2x-1)+0.188C_{2}^{3/2}(2x-1)-0.13C_{4}^{3/2}(2x-1)\right],
$$
$$
\Phi_{CLEO}(x,\mu_{0}^2)=\Phi_{asy}(x)\left[C_{0}^{3/2}(2x-1)+0.27C_{2}^{3/2}(2x-1)-0.22C_{4}^{3/2}(2x-1)\right],
$$
\begin{equation}
\Phi_{BF}(x,\mu_{0}^2)=\Phi_{asy}(x)\left[C_{0}^{3/2}(2x-1)+0.44C_{2}^{3/2}(2x-1)+0.25C_{4}^{3/2}(2x-1)\right],
\end{equation}
$$
C_{0}^{3/2}(2x-1)=1,\,\,C_{2}^{3/2}(2x-1)=\frac{3}{2}(5(2x-1)^2-1),
$$
$$
C_{4}^{3/2}(2x-1)=\frac{15}{8}(21(2x-1)^4-14(2x-1)^2+1).
$$

where $f_{\pi}=0.923 GeV$, $f_{\rho}^L$=0.141 GeV$,
f_{\rho}^{T}$=0.16 GeV are the pion and $\rho$ mesons  decay
constants. Here, we have denoted by $x\equiv x_1$, the longitudinal
fractional momentum carried by the quark within the meson. Then,
$x_2=1-x$ and $x_1-x_2=2x-1$. The pion  and $\rho$ meson wave
functions is symmetric under the replacement $x_1-x_2\leftrightarrow
x_2-x_1$.

Several important nonperturbative tools have been developed which
allow specific predictions for the hadronic wave functions directly
from theory and experiments. The QCD sum-rule technique and lattice
gauge theory provide constraints on the moments of the hadronic
distribution amplitude. However, the correct meson wave
function is still an open problem in QCD. It is known that the meson
wave function can be expanded over the eigenfunctions of the
one-loop Brodsky-Lepage equation, \emph{i.e.}, in terms of the
Gegenbauer polynomials $\{C_{n}^{3/2}(2x-1)\},$

\begin{equation}
\Phi_{M}(x,Q^2)=\Phi_{asy}(x)\left[1+\sum_{n=2..}^{\infty}a_{n}(Q^2)C_{n}^{3/2}(2x-1)\right],
\end{equation}

In the present work, we take into account the evolution of the meson
wave function on the factorization scale. The evolution of the wave
function on the factorization scale $Q^2$ is governed by the
functions $a_n(Q^2)$,

In the case $\pi$ meson,

\begin {equation}
a_n(Q^2)=a_n(\mu_{0}^2)\left[\frac{\alpha_{s}(Q^2)}{\alpha_{s}(\mu_{0}^2)}\right]^{\gamma_n/\beta_0},
\end{equation}

$$
\frac{\gamma_2}{\beta_{0}}=\frac{50}{81},\,\,\,\frac{\gamma_4}{\beta_{0}}=\frac{364}{405},\,\,
n_f=3.
$$
 In Eq.(2.6), $\{\gamma_n\}$ are anomalous dimensions defined by
the expression,

\begin{equation}
\gamma_n=C_F\left[1-\frac{2}{(n+1)(n+2)}+4\sum_{j=2}^{n+1}
\frac{1}{j}\right].
\end{equation}

In the case $\rho$ meson,

\begin {equation}
a_n(Q^2)^{\|(\bot)}=a_n(\mu_{0}^2)^{\|(\bot)}\left[\frac{\alpha_{s}(Q^2)}{\alpha_{s}(\mu_{0}^2)}\right]^{(\gamma_n^{\|(\bot)}-\gamma_0)/(2\beta_0)},
\end{equation}
Here,
$$
\gamma_n^{\|}=\frac{8}{3}\left[1-\frac{2}{(n+1)(n+2)}+4\sum_{j=2}^{n+1}
\frac{1}{j}\right].
$$
$$
\gamma_n^{\bot}=\frac{8}{3}\left[1+4\sum_{j=2}^{n+1}
\frac{1}{j}\right].
$$

The constants $a_n(\mu_{0}^2)=a_{n}^0$ are input parameters that
form the shape of the wave functions and which can be extracted from
experimental data or obtained from the nonperturbative QCD
computations at the normalization point $\mu_{0}^2$. The QCD
coupling constant $\alpha_{s}(Q^2)$ at the one-loop approximation is
given by the expression

\begin{equation}
\alpha_{s}(Q^2)=\frac{4\pi}{\beta_0 ln(Q^2/\Lambda^2)}.
\end{equation}

Here, $\Lambda$ is the fundamental QCD scale parameter, $\beta_0$ is
the QCD beta function one-loop coefficient,
$$
\beta_0=11-\frac{2}{3}n_f.
$$

The higher-twist subprocess $\gamma q\to Mq$ contributes to
$\gamma\gamma \to MX$ through the diagram of Fig.1(a). We now incorporate the
higher-twist(HT) subprocess $\gamma q\to Mq$ into the full inclusive cross
section. In this subprocess $\gamma q\to Mq$, photon and the meson may be
viewed as an effective current striking the incoming quark line. With this in
mind, we write the complete cross section in formal analogy with
deep-inelastic scattering,

\begin{equation}
E\frac{d\sigma}{d^{3}p}(\gamma \gamma\to MX )=\frac
{3}{\pi}\sum_{q \overline{q}}\int_{0}^{1}dx \delta(\hat s+\hat
t+\hat u)\hat s G_{q/{\gamma}}(x,-\hat t)\frac{d\sigma}{d\hat
t}(\gamma q \to Mq)+ (t\leftrightarrow u),
\end{equation}

Here $G_{q/\gamma}$ is the per color distribution function for a
quark in a photon. The subprocess cross section for $\pi,\rho_{L}$
and $\rho_{T}$ production

\begin{equation}
\frac{d\sigma}{d\hat t}(\gamma q\to Mq)=\left\{\begin{array}{cc}
\frac{8\pi^2\alpha_{E}C_{F}}{9}[D(\hat s,\hat u)]^{2}
\frac{1}{\hat{s}^2(-\hat t)}\left[\frac{1}{\hat{s}^2}+
\frac{1}{\hat{u}^2}\right],\,\,\, M=\pi,\rho_{L},\\
\frac{8\pi^2\alpha_{E}C_{F}}{9} \left[D(\hat s,\hat
u)\right]^2\frac{8(-\hat t)}{\hat{s}^4 \hat{u}^2},M=\rho_{T},
\end{array} \right.
\end{equation}

 where

\begin{equation}
D(\hat s,\hat u)=e_1\hat
u\int_{0}^{1}dx_1\left[\frac{\alpha_{s}(Q_1^2)\Phi_{M}(x_1,Q_1^2)}{x_1(1-x_1)}\right]+e_2\hat
s\int_{0}^{1}dx_1\left[\frac{\alpha_{s}(Q_2^2)\Phi_{M}(x_1,Q_2^2)}{x_1(1-x_1)}\right].
\end{equation}

where $Q_{1}^2=\hat
s/2,\,\,\,\,Q_{2}^2=-\hat u/2$,\,\, represents the momentum squared
carried by the hard gluon in Fig.1(a), $e_1(e_2)$ is the charge of
$q_1(\overline{q}_2)$ and $C_F=\frac{4}{3}$.

In the running coupling method this cross section was found in
Ref.[52]. For pseudoscalar and longitudinally polarized  meson
$$
\frac{d{\hat \sigma}^{HT}(e_1,e_2)}{d{\hat
t}}=\frac{32\pi^2 C_{F}\alpha_{E}}{9{\hat s}^2}\left
[-{\frac{e_{1}^2}{{\hat s}^2}}[I_{1}^2\hat t-2I_{1}(I_1\hat
s+I_{2}\hat u)\frac{\hat u}{\hat t}+I_{2}^2\frac{{\hat u}^2}{\hat
t}]-\frac{e_{2}^2}{{\hat u}^2}[K_{1}^2\hat t-2K_{1}(K_1\hat
u+K_{2}\hat s)\frac{\hat s}{\hat t}+ \right.
$$
\begin{equation}
\left. K_{2}^2\frac{{\hat s}^2}{\hat
t}]-\frac{2e_{1}e_{2}}{\hat s \hat u \hat t}[I_{1}K_{1}{\hat
t}^2-I_{1}(K_{2}\hat s+K_{1}\hat u)\hat s-K_{1}(I_1\hat s+I_{2}\hat
u)\hat u]\right].
\end{equation}
for the transversely polarized vector meson,
\begin{equation}
\frac{d{\hat \sigma}^{HT}(e_1,e_2)}{d{\hat
t}}=\frac{64\pi^2C_{F}\alpha_{E}}{9{\hat s}^4}\frac{-\hat t}{{\hat
u}^2}[e_{1}\hat u I_2-e_{2}\hat s K_2]^2
\end{equation}

Here,
$$
I_1(\hat
s)=\int_{0}^{1}\int_{0}^{1}\frac{dx_{1}dx_{2}\delta(1-x_1-x_2)\alpha_{s}((1-x_1)\hat
s)\Phi_{M}(x,Q^2)}{x_2}
$$
$$
I_2(\hat
s)=\int_{0}^{1}\int_{0}^{1}\frac{dx_{1}dx_{2}\delta(1-x_1-x_2)\alpha_{s}((1-x_1)\hat
s)\Phi_{M}(x,Q^2)}{x_{1}x_{2}}
$$
and
$$
K_1(\hat
u)=\int_{0}^{1}\int_{0}^{1}\frac{dx_{1}dx_{2}\delta(1-x_1-x_2)\alpha_{s}(-x_1\hat
u)\Phi_{M}(x,Q^2)}{x_1}
$$
$$
K_2(\hat
u)=\int_{0}^{1}\int_{0}^{1}\frac{dx_{1}dx_{2}\delta(1-x_1-x_2)\alpha_{s}(-x_1\hat
u))\Phi_{M}(x,Q^2)}{x_{1}x_{2}}
$$

The full cross section for $\pi$ and $\rho_{L}$ production is
given by
$$
E\frac{d\sigma}{d^{3}p}(\gamma \gamma\to MX )=\frac{s}{s+u}
\sum_{q\overline{q}}G_{q/{\gamma}}(x,-\hat
t)\frac{8\pi\alpha_{E}C_{F}}{3}\frac{[D(\hat s,\hat u)]^2}{{\hat
s}^2(-\hat t)}\left[\frac{1}{{\hat s}^2}+\frac{1}{{\hat
u}^2}\right]+
$$
\begin{equation}
\frac{s}{s+t} \sum_{q\overline{q}}G_{q/{\gamma}}(x,-\hat
u)\frac{8\pi\alpha_{E}C_{F}}{3}\frac{[D(\hat s,\hat t)]^2}{{\hat
s}^2(-\hat u)}\left[\frac{1}{{\hat s}^2}+\frac{1}{{\hat
t}^2}\right] ,
\end{equation}

In (2.15), the subprocess invariants are
\begin{equation}
\hat s=xs,\,\,\,\,\hat u=xu,\,\,\,\,\hat t=t,
\end{equation}

$$t=-\frac{s}{2}(x_R-x_F)= -m_T \sqrt{s} e^{-y},$$
$$u=-\frac{s}{2}(x_R+x_F)= -m_T \sqrt {s} e^y,$$
with $x_R=(x_{F}^2+x_{T}^2)^{1/2}$. Here
$x_F=2(p_M)_{\parallel}/\sqrt s$ and $x_T=2(p_M)_{\perp}/\sqrt
s=2p_{T}/\sqrt s$ specify the longitudinal and transverse momentum
of the meson. In terms of these the rapidity of $M$ is given by

$$
y=\frac{1}{2}[(x_R+x_F)/(x_R-x_F)]
$$

where $m_T$ is the transverse mass of meson, which is given by
$$m_T=\sqrt{m^2+p_{T}^2}$$
As seen from (2.11) the subprocess cross section for longitudinal
$\rho_{L}$ production is very similar to that for $\pi$ production,
but the transverse $\rho_{T}$ subprocess cross section  has a quite
different form.

Let as first consider the frozen coupling approach. In this
approach we take equal the four-momentum  square $\hat{Q}_{1,2}^2$
of the hard gluon to the meson's transverse momentum square
$\hat{Q}_{1,2}^2=p_{T}^2$. In this case the QCD coupling constant
$\alpha_s$ in the integral (2.12) does not depend on integration
variable. After this substitution, calculation of integral (2.12)
becomes easy. Hence, the effective cross section obtained after
substitution of the integral (2.12) into the expression (2.15) is
referred as the frozen coupling effective cross section. We will
denote the higher-twist cross section obtained using the frozen
coupling constant approximation by $(\Sigma_{M}^{HT})^0$.

For a full discussion,  we consider a difference $\Delta^{HT}$
between the higher-twist cross section combinations
$\Sigma_{M^{+}}^{HT}$ and $\Sigma_{M^{-}}^{HT}$

\begin{equation}
\Delta_{M}^{HT}=\Sigma_{M^{+}}^{HT}
-\Sigma_{M^{-}}^{HT}=E_{{M}^{+}}\frac{d\sigma}{d^3p}
(\gamma\gamma \to M^{+}X)-E_{{M}^{-}}\frac{d\sigma}{d^3p}(\gamma\gamma \to
M^{-}X).
\end{equation}

We have extracted the following higher-twist subprocesses
contributing to the two covariant cross sections in Eq.(2.11)

\begin{equation}
\gamma q_{1}\to(q_1\overline{q}_2)q_2 \,\,\,\,\,,
\gamma\overline{q}_{2}\to(q_1\overline{q}_2)\overline{q}_2
\end{equation}

As seen from Eq.(2.15), at fixed $p_T$, the cross section falls
very slowly with $s$. Also, at fixed $s$, the cross section
decreases as $1/p_{T}^5$, multiplied by a slowly varying
logarithmic function which vanishes at the phase-spase boundary.
Thus, the $p_T$ spectrum is  fairly independent of $s$ expect near
the kinematic limit.

\section{THE RUNNING COUPLING APPROACH AND HIGHER-TWIST MECHANISM}\label{ir}
In this section we shall calculate the integral (2.12) using the
running coupling constant  approach and also discuss the problem
of normalization of the higher-twist process cross section in the
context of the same approach.

As  is seen from (2.12), in general, one has to take into account
not only the dependence of $\alpha(\hat {Q}_{1,2}^2)$ on the scale
$\hat {Q}_{1,2}^2$, but also an evolution of $\Phi(x,\hat
{Q}_{1,2}^2)$ with $\hat {Q}_{1,2}^2$. The meson wave function
evolves in accordance with a Bethe-Salpeter-type equation.
Therefore, it is worth noting that, the renormalization scale
(argument of $\alpha_s$) should be equal to $Q_{1}^2=x_2\hat s$,
$Q_{2}^2=-x_1\hat u$, whereas the factorization scale [$Q^2$ in
$\Phi_{M}(x,Q^2)$] is taken independent from $x$, we take
$Q^2=p_{T}^2$. Such approximation does not considerably change the
numerical results, but the phenomenon considered in this article
(effect of infrared renormalons) becomes transparent. The main
problem in our investigation is the calculation of the integral in
(2.12) by the running coupling constant approach. The integral in
Eq.(2.12) in the framework of the running coupling approach takes
the form

\begin{equation}
I(\mu_{R_{0}}^2)=\int_{0}^{1}\frac{\alpha_{s}(\lambda
\mu_{R_0}^2)\Phi_{M}(x,\mu_{F}^2)dx}{1-x}.
\end{equation}

The $\alpha_{s}(\lambda \mu_{R_0}^2)$ has the infrared singularity
at $x\rightarrow1$, if $\lambda=1-x$  or $x\rightarrow0$, if
$\lambda=x$ and  as a result integral $(3.1)$ diverges (the pole
associated with the denominator of the integrand is fictitious,
because $\Phi_{M}\sim(1-x)$, and therefore, the singularity of the
integrand at $x=1$  is caused only by
$\alpha_{s}((1-x)\mu_{R_0}^2)$). For the regularization of the
integral we express the running coupling at scaling variable
$\alpha_{s}(\lambda \mu_{R_0}^2)$ with the aid of the
renormalization group equation in terms of the fixed one
$\alpha_{s}(Q^2)$. The renormalization group equation for the
running coupling $\alpha\equiv\alpha_{s}/\pi$ has the form [31]

\begin{equation}
\frac{\partial\alpha(\lambda Q^2)}{\partial
ln\lambda}\simeq-\frac{\beta_{0}}{4}[\alpha(\lambda Q^2)]^2
\end{equation}
where
$$
\beta_{0}=11-\frac{2}{3}n_{f}.
$$
The solution of Eq.(3.2), with the initial condition
$$
\alpha(\lambda)|_{\lambda=1}=\alpha\equiv\alpha_{s}(Q^2)/\pi,
$$
is [31]
\begin{equation}
\frac{\alpha(\lambda)}{\alpha}=\left[1+\alpha\frac{\beta_{0}}{4}ln{\lambda}\right]^{-1}
\end{equation}
This transcendental equation can be solved iteratively by keeping
the leading $\alpha^kln^k\lambda$ order. This term is given by
\begin{equation}
\alpha_{s}(\lambda Q^2)\simeq\frac{\alpha_{s}(Q^2)}{1+ln\lambda/t}
\end{equation}
After substituting Eq.(3.4) into Eq.(2.12) we get
$$
D(\hat s,\hat u)=e_{1}\hat t\int_{0}^{1}dx\frac{\alpha_{s}(\lambda
\mu_{R_0}^2)\Phi_{M}(x,Q^2)}{x(1-x)}+ e_{2}\hat
u\int_{0}^{1}dx\frac{\alpha_{s}(\lambda
\mu_{R_0}^2)\Phi_{M}(x,Q^2)}{x(1-x)}=
$$
$$
e_{1}\hat t\alpha_{s}(\hat s)\int_{0}^{1}dx
\frac{\Phi_{M}(x,Q^2)}{x(1-x)(1+ln\lambda/t_{1})}+ e_{2}\hat u
\alpha_{s}(-\hat
u)\int_{0}^{1}dx\frac{\Phi_{M}(x,Q^2)}{x(1-x)(1+ln\lambda/t_{2})}=
$$
$$
e_{1}\hat t\alpha_{s}(\hat s)\int_{0}^{1}dx
\frac{\Phi_{asy}(x)\left[1+\sum_{2,4,..}^{\infty}a_{n}(\mu_{0}^{2})
\left[\frac{\alpha_{s}(Q^2)}
{\alpha_{s}(\mu_{0}^2)}\right]^{\gamma_{n}/\beta_{0}}C_{n}^{3/2}(2x-1)\right]}{x(1-x)
(1+ln\lambda/t_{1})}+
$$
\begin{equation}
e_{2}\hat u\alpha_{s}(-\hat u)\int_{0}^{1}dx
\frac{\Phi_{asy}(x)\left[1+\sum_{2,4,..}^{\infty}a_{n}(\mu_{0}^{2})
\left[\frac{\alpha_{s}(Q^2)}{\alpha_{s}(\mu_{0}^2)}\right]^{\gamma_{n}/\beta_{0}}C_{n}^{3/2}(2x-1)\right]}{x(1-x)(1+ln\lambda/t_{2})},
\end{equation}
where $t_1=4\pi/\alpha_{s}(\hat s)\beta_{0},
t_2=4\pi/\alpha_{s}(-\hat u)\beta_{0}$. The integral (3.5) is
common and, of course, still divergent, but now it is recast into
a form, which is suitable for calculation. Using the running
coupling constant approach, this integral may be found as a
perturbative series in $\alpha_{s}$
\begin{equation}
D(\hat s,\hat u)\sim
\sum_{n=1}^{\infty}\left(\frac{\alpha_{s}}{4\pi}\right)^nS_{n}.
\end{equation}
The expression coefficients $S_n$ can be written as power series
in the number of light quark flavors or, equivalently, as a series
in power of $\beta_0$, as $S_{n}=C_{n}\beta_{0}^{n-1}$. The
coefficients $C_n$ of this series demonstrate factorial growth
$C_n\sim(n-1)!$, which might indicate an infrared renormalon
nature of divergences in the integral (3.5) and corresponding
series (3.6). The procedure for dealing with such ill-defined
series is well known; one has to perform the Borel transform of
the series [53]
$$
B[D](u)=\sum_{n=0}^{\infty}\frac{D_n}{n!} u^n,
$$
then invert $B[D](u)$ to obtain the resummed expression (the Borel
sum) $D(\hat s,\hat u)$. After this we can find directly the
resummed expression for $D(\hat s,\hat u)$. The change of the
variable $x$ to $z=ln(1-x)$, as $ln(1-x)=ln\lambda$. Then,
\begin{equation}
D(\hat s,\hat u)=e_{1}\hat t \alpha_{s}(\hat s) t_{1} \int_{0}^{1}
\frac{\Phi_{M}(x,Q^2)dx}{x(1-x)(t_1+z)}+
e_{2}\hat u \alpha_{s}(-\hat u) t_2 \int_{0}^{1}
\frac{\Phi_{M}(x,Q^2)dx}{x(1-x)(t_2+z)}
\end{equation}

For the calculation the expression (3.7) we will apply the
integral representation of $1/(t+z)$ [54]. After this operation,
formula (3.7) is simplified and we can extract the Borel sum of
the perturbative series (3.6) and the corresponding Borel
transform in dependence from the wave functions of the meson,
respectively. Also after such manipulations the obtained
expression can be used for numerical computations.

It is convenient to use  the following integral representation for $1/(t+z)$:
\begin{equation}
\frac{1}{t+z}=\int_{0}^{\infty}e^{-(t+z)u}du
\end{equation}
After inserting Eq.(3.8) into (3.7),
 then, we obtain
$$
D(\hat s,\hat u)=e_{1} \hat{t} \alpha_{s}(\hat s) t_1 \int_{0}^{1}
\int_{0}^{\infty} \frac{\Phi_{M}(x,Q^2)e^{-(t_1+z)u}du
dx}{x(1-x)}+
$$
\begin{equation}
e_{2} \hat{u} \alpha_{s}(-\hat u) t_2 \int_{0}^{1}
\int_{0}^{\infty} \frac{\Phi_{M}(x,Q^2)e^{-(t_2+z)u}du
dx}{x(1-x)}.
\end{equation}
In the case of $\Phi_{asy}(x)$ for $I_1(\hat s),I_2(\hat
s),K_1(\hat u),K_2(\hat u)$, we get
$$
I_1(\hat s)=\frac{4\sqrt{3} \pi f_{\pi}}{\beta_{0}}
\int_{0}^{\infty}du e^{-t_{1}u}\left[\frac{1}{1-u}-
\frac{1}{2-u}\right].
$$
$$
I_2(\hat s)=\frac{4\sqrt{3} \pi f_{\pi}}{\beta_{0}}
\int_{0}^{\infty}du e^{-t_{1}u}\left[\frac{1}{1-u}\right].
$$
$$
K_1(\hat u)=\frac{4\sqrt{3} \pi f_{\pi}}{\beta_{0}}
\int_{0}^{\infty}du e^{-t_{2}u}\left[\frac{1}{1-u}-
\frac{1}{2-u}\right].
$$
\begin{equation}
K_2(\hat u)=\frac{4\sqrt{3} \pi f_{\pi}}{\beta_{0}}
\int_{0}^{\infty}du e^{-t_{2}u}\left[\frac{1}{1-u}\right].
\end{equation}
In the case  of the $\Phi_{L(T)}^{\rho}(x,Q^2)$ wave function, we find
$$
I_1(\hat s)=\frac{4\sqrt{6}\pi f_{\rho}}{\beta_{0}}
\int_{0}^{\infty}du e^{-t_{1}u}
\left[\frac{1}{1-u}-\frac{1}{2-u}+\right.
$$
$$
0.27(0.3)\left[\frac{\alpha_{s}(Q^2)}{\alpha_{s}(\mu_{0}^2)}\right]^{(50/162),(26/162)}
\left[\frac{4}{1-u}-\frac{24}{2-u}+\frac{40}{3-u}-
\left.\frac{20}{4-u}\right]\right],
$$
$$
I_2(\hat s)=\frac{4\sqrt{6}\pi f_{\rho}}{\beta_{0}}
\int_{0}^{\infty}du e^{-t_{1}u} \left[\frac{1}{1-u}+\right.
$$
$$
0.27(0.3)\left[\frac{\alpha_{s}(Q^2)}{\alpha_{s}(\mu_{0}^2)}\right]^{(50/162),(26/162)}
\left[\frac{4}{1-u}-\frac{20}{2-u}+\left.\frac{20}{3-u}
\right]\right],
$$
$$
K_1(\hat u)=\frac{4\sqrt{6}\pi f_{\rho}}{\beta_{0}}
\int_{0}^{\infty}du e^{-t_{2}u}
\left[\frac{1}{1-u}-\frac{1}{2-u}+\right.
$$
$$
0.27(0.3)\left[\frac{\alpha_{s}(Q^2)}{\alpha_{s}(\mu_{0}^2)}\right]^{(50/162),(26/162)}
\left[\frac{4}{1-u}-\frac{24}{2-u}+\frac{40}{3-u}-
\left.\frac{20}{4-u}\right]\right],
$$
$$
K_2(\hat u)=\frac{4\sqrt{6}\pi f_{\rho}}{\beta_{0}}
\int_{0}^{\infty}du e^{-t_{2}u} \left[\frac{1}{1-u}+\right.
$$
\begin{equation}
0.27(0.3)\left[\frac{\alpha_{s}(Q^2)}{\alpha_{s}(\mu_{0}^2)}\right]^{(50/162),(26/162)}
\left[\frac{4}{1-u}-\frac{20}{2-u}+\left.\frac{20}{3-u}
\right]\right],
\end{equation}

In the case of the $\Phi_{CLEO}(x,Q^2)$ wave function, we get
$$
I_1(\hat s)=\frac{4\sqrt{3}\pi f_{\pi}}{\beta_{0}}
\int_{0}^{\infty}du e^{-t_{1}u}\left[\frac{1}{1-u}-\frac{1}{2-u}+
\right.
0.405\left[\frac{\alpha_{s}(Q^2)}{\alpha_{s}(\mu_{0}^2)}\right]^{50/81}\cdot
$$
$$
\left[\frac{4}{1-u}-\frac{24}{2-u}+\frac{40}{3-u}-
\frac{20}{4-u}\right]-0.4125\left[\frac{\alpha_{s}(Q^2)}{\alpha_{s}(\mu_{0}^2)}\right]^{364/405}\cdot
$$
$$
\left.
\left[\frac{8}{1-u}-\frac{120}{2-u}+\frac{560}{3-u}-\frac{1112}{4-u}+\frac{1008}{5-u}-\frac{336}{6-u}\right]\right].
$$
$$
I_2(\hat s)=\frac{4\sqrt{3}\pi f_{\pi}}{\beta_{0}}
\int_{0}^{\infty}du e^{-t_{1}u}\left[\frac{1}{1-u}+ \right.
0.405\left[\frac{\alpha_{s}(Q^2)}{\alpha_{s}(\mu_{0}^2)}\right]^{50/81}\cdot
$$
$$
\left[\frac{4}{1-u}-\frac{20}{2-u}+\frac{20}{3-u}
\right]-0.4125\left[\frac{\alpha_{s}(Q^2)}{\alpha_{s}(\mu_{0}^2)}\right]^{364/405}\cdot
$$
\begin{equation}
\left.
\left[\frac{8}{1-u}-\frac{112}{2-u}+\frac{448}{3-u}-\frac{672}{4-u}+\frac{336}{5-u}\right]\right].
\end{equation}
$$
K_1(\hat u)=\frac{4\sqrt{3}\pi f_{\pi}}{\beta_{0}}
\int_{0}^{\infty}du e^{-t_{2}u}\left[\frac{1}{1-u}-\frac{1}{2-u}+
\right.
0.405\left[\frac{\alpha_{s}(Q^2)}{\alpha_{s}(\mu_{0}^2)}\right]^{50/81}\cdot
$$
$$
\left[\frac{4}{1-u}-\frac{24}{2-u}+\frac{40}{3-u}-
\frac{20}{4-u}\right]-0.4125\left[\frac{\alpha_{s}(Q^2)}{\alpha_{s}(\mu_{0}^2)}\right]^{364/405}\cdot
$$
$$
\left.
\left[\frac{8}{1-u}-\frac{120}{2-u}+\frac{560}{3-u}-\frac{1112}{4-u}+\frac{1008}{5-u}-\frac{336}{6-u}\right]\right].
$$
$$
K_2(\hat u)=\frac{4\sqrt{3}\pi f_{\pi}}{\beta_{0}}
\int_{0}^{\infty}du e^{-t_{2}u}\left[\frac{1}{1-u}+ \right.
0.405\left[\frac{\alpha_{s}(Q^2)}{\alpha_{s}(\mu_{0}^2)}\right]^{50/81}\cdot
$$
$$
\left[\frac{4}{1-u}-\frac{20}{2-u}+\frac{20}{3-u}
\right]-0.4125\left[\frac{\alpha_{s}(Q^2)}{\alpha_{s}(\mu_{0}^2)}\right]^{364/405}\cdot
$$
$$
\left.
\left[\frac{8}{1-u}-\frac{112}{2-u}+\frac{448}{3-u}-\frac{672}{4-u}+\frac{336}{5-u}\right]\right].
$$
Also, in the case of the $\Phi_{BMS}(x,Q^2)$ wave function, we get
$$
I_1(\hat s)=\frac{4\sqrt{3}\pi f_{\pi}}{\beta_{0}}
\int_{0}^{\infty}du e^{-t_{1}u}\left[\frac{1}{1-u}-\frac{1}{2-u}+
\right.
0.282\left[\frac{\alpha_{s}(Q^2)}{\alpha_{s}(\mu_{0}^2)}\right]^{50/81}\cdot
$$
$$
\left[\frac{4}{1-u}-\frac{24}{2-u}+\frac{40}{3-u}-
\frac{20}{4-u}\right]-0.244\left[\frac{\alpha_{s}(Q^2)}{\alpha_{s}(\mu_{0}^2)}\right]^{364/405}\cdot
$$
$$
\left.
\left[\frac{8}{1-u}-\frac{120}{2-u}+\frac{560}{3-u}-\frac{1112}{4-u}+\frac{1008}{5-u}-\frac{336}{6-u}\right]\right].
$$
$$
I_2(\hat s)=\frac{4\sqrt{3}\pi f_{\pi}}{\beta_{0}}
\int_{0}^{\infty}du e^{-t_{1}u}\left[\frac{1}{1-u}+ \right.
0.282\left[\frac{\alpha_{s}(Q^2)}{\alpha_{s}(\mu_{0}^2)}\right]^{50/81}\cdot
$$
$$
\left[\frac{4}{1-u}-\frac{20}{2-u}+\frac{20}{3-u}
\right]-0.244\left[\frac{\alpha_{s}(Q^2)}{\alpha_{s}(\mu_{0}^2)}\right]^{364/405}\cdot
$$
$$
\left.
\left[\frac{8}{1-u}-\frac{112}{2-u}+\frac{448}{3-u}-\frac{672}{4-u}+\frac{336}{5-u}\right]\right].
$$
$$
K_1(\hat u)=\frac{4\sqrt{3}\pi f_{\pi}}{\beta_{0}}
\int_{0}^{\infty}du e^{-t_{2}u}\left[\frac{1}{1-u}-\frac{1}{2-u}+
\right.
0.282\left[\frac{\alpha_{s}(Q^2)}{\alpha_{s}(\mu_{0}^2)}\right]^{50/81}\cdot
$$
$$
\left[\frac{4}{1-u}-\frac{24}{2-u}+\frac{40}{3-u}-
\frac{20}{4-u}\right]-0.244\left[\frac{\alpha_{s}(Q^2)}{\alpha_{s}(\mu_{0}^2)}\right]^{364/405}\cdot
$$
$$
\left.
\left[\frac{8}{1-u}-\frac{120}{2-u}+\frac{560}{3-u}-\frac{1112}{4-u}+\frac{1008}{5-u}-\frac{336}{6-u}\right]\right].
$$
$$
K_2(\hat u)=\frac{4\sqrt{3}\pi f_{\pi}}{\beta_{0}}
\int_{0}^{\infty}du e^{-t_{2}u}\left[\frac{1}{1-u}+ \right.
0.282\left[\frac{\alpha_{s}(Q^2)}{\alpha_{s}(\mu_{0}^2)}\right]^{50/81}\cdot
$$
$$
\left[\frac{4}{1-u}-\frac{20}{2-u}+\frac{20}{3-u}
\right]-0.244\left[\frac{\alpha_{s}(Q^2)}{\alpha_{s}(\mu_{0}^2)}\right]^{364/405}\cdot
$$
\begin{equation}
\left.
\left[\frac{8}{1-u}-\frac{112}{2-u}+\frac{448}{3-u}-\frac{672}{4-u}+\frac{336}{5-u}\right]\right].
\end{equation}

Equation(3.1) and (3.2) is nothing more than the Borel sum of the
perturbative series (3.6), and the corresponding Borel transform in
the case $\Phi_{asy}(x)$ is
$$
B[I_1](u)=\frac{1}{1-u}-\frac{1}{2-u},
$$
$$
B[I_2](u)=\frac{1}{1-u},
$$
$$
B[K_1](u)=\frac{1}{1-u}-\frac{1}{2-u},
$$
\begin{equation}
B[K_2](u)=\frac{1}{1-u},
\end{equation}
in the case $\Phi_{L(T)}^{\rho}(x,Q^2)$ is
$$
B[I_1](u)=\frac{1}{1-u}-\frac{1}{2-u}+0.27(0.3)\left(\frac{\alpha_{s}(Q^2)}{\alpha_{s}(\mu_{0}^2)}\right)^{(50/162),(26/162)}
\left(\frac{4}{1-u}-\frac{24}{2-u}+\frac{40}{3-u}-\frac{20}{4-u}\right),
$$
$$
B[I_2](u)=\frac{1}{1-u}+0.27(0.3)\left(\frac{\alpha_{s}(Q^2)}{\alpha_{s}(\mu_{0}^2)}\right)^{(50/162),(26/162)}
\left(\frac{4}{1-u}-\frac{20}{2-u}+\frac{20}{3-u}\right),
$$
$$
B[K_1](u)=\frac{1}{1-u}-\frac{1}{2-u}+0.27(0.3)\left(\frac{\alpha_{s}(Q^2)}{\alpha_{s}(\mu_{0}^2)}\right)^{(50/162),(26/162)}
\left(\frac{4}{1-u}-\frac{24}{2-u}+\frac{40}{3-u}-\frac{20}{4-u}\right),
$$
\begin{equation}
B[K_2](u)=\frac{1}{1-u}+0.27(0.3)\left(\frac{\alpha_{s}(Q^2)}{\alpha_{s}(\mu_{0}^2)}\right)^{(50/162),(26/162)}
\left(\frac{4}{1-u}-\frac{20}{2-u}+\frac{20}{3-u}\right),
\end{equation}

in the case $\Phi_{CLEO}(x,Q^2)$ is
$$
B[I_1](u)=\frac{1}{1-u}-\frac{1}{2-u}+0.405\left(\frac{\alpha_{s}(Q^2)}{\alpha_{s}(\mu_{0}^2)}\right)^{50/81}
\left(\frac{4}{1-u}-\frac{24}{2-u}+\frac{40}{3-u}-\frac{20}{4-u}\right)-
$$
$$
0.4125\left(\frac{\alpha_{s}(Q^2)}{\alpha_{s}(\mu_{0}^2)}\right)^{364/405}\left(\frac{8}{1-u}-
\frac{120}{2-u}+\frac{560}{3-u}-\frac{1112}{4-u}+\frac{1008}{5-u}-\frac{336}{6-u}\right).
$$
$$
B[I_2](u)=\frac{1}{1-u}+0.405\left(\frac{\alpha_{s}(Q^2)}{\alpha_{s}(\mu_{0}^2)}\right)^{50/81}
\left(\frac{4}{1-u}-\frac{20}{2-u}+\frac{20}{3-u}\right)-
$$
$$
0.4125\left(\frac{\alpha_{s}(Q^2)}{\alpha_{s}(\mu_{0}^2)}\right)^{364/405}\left(\frac{8}{1-u}-
\frac{112}{2-u}+\frac{448}{3-u}-\frac{672}{4-u}+\frac{336}{5-u}\right).
$$
$$
B[K_1](u)=\frac{1}{1-u}-\frac{1}{2-u}+0.405\left(\frac{\alpha_{s}(Q^2)}{\alpha_{s}(\mu_{0}^2)}\right)^{50/81}
\left(\frac{4}{1-u}-\frac{24}{2-u}+\frac{40}{3-u}-\frac{20}{4-u}\right)-
$$
$$
0.4125\left(\frac{\alpha_{s}(Q^2)}{\alpha_{s}(\mu_{0}^2)}\right)^{364/405}\left(\frac{8}{1-u}-
\frac{120}{2-u}+\frac{560}{3-u}-\frac{1112}{4-u}+\frac{1008}{5-u}-\frac{336}{6-u}\right).
$$
$$
B[K_2](u)=\frac{1}{1-u}+0.405\left(\frac{\alpha_{s}(Q^2)}{\alpha_{s}(\mu_{0}^2)}\right)^{50/81}
\left(\frac{4}{1-u}-\frac{20}{2-u}+\frac{20}{3-u}\right)-
$$
\begin{equation}
0.4125\left(\frac{\alpha_{s}(Q^2)}{\alpha_{s}(\mu_{0}^2)}\right)^{364/405}\left(\frac{8}{1-u}-
\frac{112}{2-u}+\frac{448}{3-u}-\frac{672}{4-u}+\frac{336}{5-u}\right).
\end{equation}
and in the case $\Phi_{BMS}(x,Q^2)$ is
$$
B[I_1](u)=\frac{1}{1-u}-\frac{1}{2-u}+0.282\left(\frac{\alpha_{s}(Q^2)}{\alpha_{s}(\mu_{0}^2)}\right)^{50/81}
\left(\frac{4}{1-u}-\frac{24}{2-u}+\frac{40}{3-u}-\frac{20}{4-u}\right)-
$$
$$
0.244\left(\frac{\alpha_{s}(Q^2)}{\alpha_{s}(\mu_{0}^2)}\right)^{364/405}\left(\frac{8}{1-u}-
\frac{120}{2-u}+\frac{560}{3-u}-\frac{1112}{4-u}+\frac{1008}{5-u}-\frac{336}{6-u}\right).
$$
$$
B[I_2](u)=\frac{1}{1-u}+0.282\left(\frac{\alpha_{s}(Q^2)}{\alpha_{s}(\mu_{0}^2)}\right)^{50/81}
\left(\frac{4}{1-u}-\frac{20}{2-u}+\frac{20}{3-u}\right)-
$$
$$
0.244\left(\frac{\alpha_{s}(Q^2)}{\alpha_{s}(\mu_{0}^2)}\right)^{364/405}\left(\frac{8}{1-u}-
\frac{112}{2-u}+\frac{448}{3-u}-\frac{672}{4-u}+\frac{336}{5-u}\right).
$$
$$
B[K_1](u)=\frac{1}{1-u}-\frac{1}{2-u}+0.282\left(\frac{\alpha_{s}(Q^2)}{\alpha_{s}(\mu_{0}^2)}\right)^{50/81}
\left(\frac{4}{1-u}-\frac{24}{2-u}+\frac{40}{3-u}-\frac{20}{4-u}\right)-
$$
$$
0.244\left(\frac{\alpha_{s}(Q^2)}{\alpha_{s}(\mu_{0}^2)}\right)^{364/405}\left(\frac{8}{1-u}-
\frac{120}{2-u}+\frac{560}{3-u}-\frac{1112}{4-u}+\frac{1008}{5-u}-\frac{336}{6-u}\right).
$$
$$
B[K_2](u)=\frac{1}{1-u}+0.282\left(\frac{\alpha_{s}(Q^2)}{\alpha_{s}(\mu_{0}^2)}\right)^{50/81}
\left(\frac{4}{1-u}-\frac{20}{2-u}+\frac{20}{3-u}\right)-
$$
\begin{equation}
0.244\left(\frac{\alpha_{s}(Q^2)}{\alpha_{s}(\mu_{0}^2)}\right)^{364/405}\left(\frac{8}{1-u}-
\frac{112}{2-u}+\frac{448}{3-u}-\frac{672}{4-u}+\frac{336}{5-u}\right).
\end{equation}

The series (3.6) can be recovered by means of the following formula:
$$
C_{n}=\left(\frac{d}{du}\right)^{n-1}B[D](u)\mid_{u=0}
$$
The Borel transform $B[D](u)$ has poles on the real $u$ axis at
$u=1;2;3;4;5;6,$ which confirms our conclusion concerning the
infrared renormalon nature of divergences in (3.6). To remove them
from Eqs.(3.10-3.20) some regularization methods have to be
applied. In this article we adopt the principal value
prescription. We obtain: in the case $\Phi_{asy}$
$$
[I_1(\hat s)]^{res}=\frac{4\sqrt{3}\pi f_{\pi}}{\beta_{0}}
\left[\frac{Li(\lambda_1)}{\lambda_1}-\frac{Li(\lambda_{1}^2)}{\lambda_{1}^2}\right],
$$
$$
[I_2(\hat s)]^{res}=\frac{4\sqrt{3}\pi f_{\pi}}{\beta_{0}}
\left[\frac{Li(\lambda_1)}{\lambda_1}\right],
$$
$$
[K_1(\hat u)]^{res}=\frac{4\sqrt{3}\pi f_{\pi}}{\beta_{0}}
\left[\frac{Li(\lambda_2)}{\lambda_2}-\frac{Li(\lambda_{2}^2)}{\lambda_{2}^2}\right],
$$
\begin{equation}
[K_2(\hat u)]^{res}=\frac{4\sqrt{3}\pi f_{\pi}}{\beta_{0}}
\left[\frac{Li(\lambda_2)}{\lambda_2}\right],
\end{equation}
in the case $\Phi_{L(T)}^{\rho}(x,Q^2)$
$$
[I_1(\hat s)]^{res}=\frac{4\sqrt{6}\pi f_{\rho}}{\beta_{0}}
\left[\left[\frac{Li(\lambda_1)}{\lambda_1}-\frac{Li(\lambda_{1}^2)}{\lambda_{1}^2}\right]+
0.27(0.3)\left(\frac{\alpha_{s}(Q^2)}{\alpha_{s}(\mu_{0}^2)}\right)^{(50/162),(26/162)}.\right.
$$
$$
\left[4\frac{Li(\lambda_1)}{\lambda_1}-24\frac{Li(\lambda_{1}^2)}{\lambda_{1}^2}+ \left.
40\frac{Li(\lambda_{1}^3)}{\lambda_{1}^3}-20\frac{Li(\lambda_{1}^4)}{\lambda_{1}^4}\right]\right],
$$
$$
[I_2(\hat s)]^{res}=\frac{4 \sqrt {6} \pi f_{\rho}}{\beta_{0}}
\left[\left[\frac{Li(\lambda_1)}{\lambda_1}\right]+
0.27(0.3)\left(\frac{\alpha_{s}(Q^2)}{\alpha_{s}(\mu_{0}^2)}\right)^{(50/162),(26/162)}.\right.
$$
$$
\left[4\frac{Li(\lambda_1)}{\lambda_1}-20\frac{Li(\lambda_{1}^2)}{\lambda_{1}^2}+ \left.
20\frac{Li(\lambda_{1}^3)}{\lambda_{1}^3}\right]\right],
$$
$$
[K_1(\hat u)]^{res}=\frac{4\sqrt{6}\pi f_{\rho}}{\beta_{0}}
\left[\left[\frac{Li(\lambda_2)}{\lambda_2}-\frac{Li(\lambda_{2}^2)}{\lambda_{2}^2}\right]+
0.27(0.3)\left(\frac{\alpha_{s}(Q^2)}{\alpha_{s}(\mu_{0}^2)}\right)^{(50/162),(26/162)}.\right.
$$
$$
\left[4\frac{Li(\lambda_2)}{\lambda_2}-24\frac{Li(\lambda_{2}^2)}{\lambda_{2}^2}+ \left.
40\frac{Li(\lambda_{2}^3)}{\lambda_{2}^3}-20\frac{Li(\lambda_{2}^4)}{\lambda_{2}^4}\right]\right],
$$
$$
[K_2(\hat u)]^{res}=\frac{4 \sqrt {6} \pi f_{\rho}}{\beta_{0}}
\left[\left[\frac{Li(\lambda_2)}{\lambda_2}\right]+
0.27(0.3)\left(\frac{\alpha_{s}(Q^2)}{\alpha_{s}(\mu_{0}^2)}\right)^{(50/162),(26/162)}.\right.
$$
\begin{equation}
\left[4\frac{Li(\lambda_2)}{\lambda_2}-20\frac{Li(\lambda_{2}^2)}{\lambda_{2}^2}+ \left.
20\frac{Li(\lambda_{2}^3)}{\lambda_{2}^3}\right]\right],
\end{equation}
 in the case $\Phi_{CLEO}(x,Q^2)$
$$
[I_1(\hat s)]^{res}=\frac{4\sqrt{3}\pi f_{\pi}}{\beta_{0}}
\left[\left(\frac{Li(\lambda_1)}{\lambda_1}-\frac{Li(\lambda_{1}^2)}{\lambda_{1}^2}\right)+\right.
0.405\left(\frac{\alpha_{s}(Q^2)}{\alpha_{s}(\mu_{0}^2)}\right)^{50/81}\left(4\frac{Li(\lambda_{1})}{\lambda_{1}}-\right.
$$
$$
24\frac{Li(\lambda_{1}^2)}{\lambda_{1}^2}+40\frac{Li(\lambda_{1}^3)}{\lambda_{1}^3}-
\left.20\frac{Li(\lambda_{1}^4)}{\lambda_{1}^4}\right)-
0.4125\left(\frac{\alpha_{s}(Q^2)}{\alpha_{s}(\mu_{0}^2)}\right)^{364/405}\left(8\frac{Li(\lambda_1)}{\lambda_1}-\right.
120\frac{Li(\lambda_{1}^2)}{\lambda_{1}^2}+560\frac{Li(\lambda_{1}^3)}{\lambda_{1}^3}-
$$
$$
1112\frac{Li(\lambda_{1}^4)}{\lambda_{1}^4}+
1008\frac{Li(\lambda_{1}^5)}{\lambda_{1}^5}-
\left.\left.336\frac{Li(\lambda_{1}^6)}{\lambda_{1}^6}\right)\right],
$$
$$
[I_2(\hat s)]^{res}=\frac{4\sqrt{3}\pi f_{\pi}}{\beta_{0}}
\left[\left(\frac{Li(\lambda_1)}{\lambda_1}\right)+\right.
0.405\left(\frac{\alpha_{s}(Q^2)}{\alpha_{s}(\mu_{0}^2)}\right)^{50/81}\left(4\frac{Li(\lambda_1)}{\lambda_1}-\right.
$$
$$
\left.
20\frac{Li(\lambda_{1}^2)}{\lambda_{1}^2}+20\frac{Li(\lambda_{1}^3)}{\lambda_{1}^3}
\right)-
0.4125\left(\frac{\alpha_{s}(Q^2)}{\alpha_{s}(\mu_{0}^2)}\right)^{364/405}\left(8\frac{Li(\lambda_1)}{\lambda_1}-\right.112\frac{Li(\lambda_{1}^2)}{\lambda_{1}^2}+448\frac{Li(\lambda_{1}^3)}{\lambda_{1}^3}-
$$
$$
\left.
672\frac{Li(\lambda_{1}^4)}{\lambda_{1}^4}+
\left.336\frac{Li(\lambda_{1}^5)}{\lambda_{1}^5}\right)\right],
$$
$$
[K_1(\hat u)]^{res}=\frac{4\sqrt{3}\pi f_{\pi}}{\beta_{0}}
\left[\left(\frac{Li(\lambda_2)}{\lambda_2}-\frac{Li(\lambda_{2}^2)}{\lambda_{2}^2}\right)+\right.
0.405\left(\frac{\alpha_{s}(Q^2)}{\alpha_{s}(\mu_{0}^2)}\right)^{50/81}\left(4\frac{Li(\lambda_2)}{\lambda_2}-\right.
$$
$$
24\frac{Li(\lambda_{2}^2)}{\lambda_{2}^2}+40\frac{Li(\lambda_{2}^3)}{\lambda_{2}^3}-
\left.20\frac{Li(\lambda_{2}^4)}{\lambda_{2}^4}\right)-
0.4125\left(\frac{\alpha_{s}(Q^2)}{\alpha_{s}(\mu_{0}^2)}\right)^{364/405}\left(8\frac{Li(\lambda_2)}{\lambda_2}-\right.
120\frac{Li(\lambda_{2}^2)}{\lambda_{2}^2}+560\frac{Li(\lambda_{2}^3)}{\lambda_{2}^3}-
$$
$$
1112\frac{Li(\lambda_{2}^4)}{\lambda_{2}^4}+
1008\frac{Li(\lambda_{2}^5)}{\lambda_{2}^5}-
\left.\left.336\frac{Li(\lambda_{2}^6)}{\lambda_{2}^6}\right)\right],
$$
$$
[K_2(\hat u)]^{res}=\frac{4\sqrt{3}\pi f_{\pi}}{\beta_{0}}
\left[\left(\frac{Li(\lambda_2)}{\lambda_2}\right)+\right.
0.405\left(\frac{\alpha_{s}(Q^2)}{\alpha_{s}(\mu_{0}^2)}\right)^{50/81}\left(4\frac{Li(\lambda_2)}{\lambda_2}-\right.
$$
$$
\left.
20\frac{Li(\lambda_{2}^2)}{\lambda_{2}^2}+20\frac{Li(\lambda_{2}^3)}{\lambda_{2}^3}
\right)-
0.4125\left(\frac{\alpha_{s}(Q^2)}{\alpha_{s}(\mu_{0}^2)}\right)^{364/405}\left(8\frac{Li(\lambda_2)}{\lambda_2}-\right.
112\frac{Li(\lambda_{2}^2)}{\lambda_{2}^2}+448\frac{Li(\lambda_{2}^3)}{\lambda_{2}^3}-
$$
\begin{equation}
672\frac{Li(\lambda_{2}^4)}{\lambda_{2}^4}+ \left.\left.
336\frac{Li(\lambda_{2}^5)}{\lambda_{2}^5}\right)\right],
\end{equation}

also in the case $\Phi_{BMS}(x,Q^2)$
$$
[I_1(\hat s)]^{res}=\frac{4\sqrt{3}\pi f_{\pi}}{\beta_{0}}
\left[\left(\frac{Li(\lambda_1)}{\lambda_1}-\frac{Li(\lambda_{1}^2)}{\lambda_{1}^2}\right)+\right.
0.282\left(\frac{\alpha_{s}(Q^2)}{\alpha_{s}(\mu_{0}^2)}\right)^{50/81}\left(4\frac{Li(\lambda_1)}{\lambda_1}-\right.
$$
$$
24\frac{Li(\lambda_{1}^2)}{\lambda_{1}^2}+40\frac{Li(\lambda_{1}^3)}{\lambda_{1}^3}-
\left.20\frac{Li(\lambda_{1}^4)}{\lambda_{1}^4}\right)-
0.244\left(\frac{\alpha_{s}(Q^2)}{\alpha_{s}(\mu_{0}^2)}\right)^{364/405}\left(8\frac{Li(\lambda_1)}{\lambda_1}-
120\frac{Li(\lambda_{1}^2)}{\lambda_{1}^2}+560\frac{Li(\lambda_{1}^3)}{\lambda_{1}^3}-
\right.
$$
$$
1112\frac{Li(\lambda_{1}^4)}{\lambda_{1}^4}+
1008\frac{Li(\lambda_{1}^5)}{\lambda_{1}^5}-
\left.\left.336\frac{Li(\lambda_{1}^6)}{\lambda_{1}^6}\right)\right],
$$
$$
[I_2(\hat s)]^{res}=\frac{4\sqrt{3}\pi f_{\pi}}{\beta_{0}}
\left[\left(\frac{Li(\lambda_1)}{\lambda_1}\right)+\right.
0.282\left(\frac{\alpha_{s}(Q^2)}{\alpha_{s}(\mu_{0}^2)}\right)^{50/81}\left(4\frac{Li(\lambda_1)}{\lambda_1}-\right.
$$
$$
\left.
20\frac{Li(\lambda_{1}^2)}{\lambda_{1}^2}+20\frac{Li(\lambda_{1}^3)}{\lambda_{1}^3}\right)-
0.244\left(\frac{\alpha_{s}(Q^2)}{\alpha_{s}(\mu_{0}^2)}\right)^{364/405}\left(8\frac{Li(\lambda_1)}{\lambda_1}-\right.
112\frac{Li(\lambda_{1}^2)}{\lambda_{1}^2}+448\frac{Li(\lambda_{1}^3)}{\lambda_{1}^3}-
$$
$$
672\frac{Li(\lambda_{1}^4)}{\lambda_{1}^4}+ \left.\left.
336\frac{Li(\lambda_{1}^5)}{\lambda_{1}^5}\right)\right],
$$
$$
[K_1(\hat u)]^{res}=\frac{4\sqrt{3}\pi f_{\pi}}{\beta_{0}}
\left[\left(\frac{Li(\lambda_2)}{\lambda_2}-\frac{Li(\lambda_{2}^2)}{\lambda_{2}^2}\right)+\right.
0.282\left(\frac{\alpha_{s}(Q^2)}{\alpha_{s}(\mu_{0}^2)}\right)^{50/81}\left(4\frac{Li(\lambda_2)}{\lambda_2}-\right.
$$
$$
24\frac{Li(\lambda_{2}^2)}{\lambda_{2}^2}+40\frac{Li(\lambda_{2}^3)}{\lambda_{2}^3}-
\left.20\frac{Li(\lambda_{2}^4)}{\lambda_{2}^4}\right)-
0.244\left(\frac{\alpha_{s}(Q^2)}{\alpha_{s}(\mu_{0}^2)}\right)^{364/405}\left(8\frac{Li(\lambda_2)}{\lambda_2}-\right.
120\frac{Li(\lambda_{2}^2)}{\lambda_{2}^2}+560\frac{Li(\lambda_{2}^3)}{\lambda_{2}^3}-
$$
$$
1112\frac{Li(\lambda_{2}^4)}{\lambda_{2}^4}+
1008\frac{Li(\lambda_{2}^5)}{\lambda_{2}^5}-
\left.\left.336\frac{Li(\lambda_{2}^6)}{\lambda_{2}^6}\right)\right],
$$
$$
[K_2(\hat u)]^{res}=\frac{4\sqrt{3}\pi f_{\pi}}{\beta_{0}}
\left[\left(\frac{Li(\lambda_2)}{\lambda_2}\right)+\right.
0.282\left(\frac{\alpha_{s}(Q^2)}{\alpha_{s}(\mu_{0}^2)}\right)^{50/81}\left(4\frac{Li(\lambda_2)}{\lambda_2}-\right.
$$
$$
20\frac{Li(\lambda_{2}^2)}{\lambda_{2}^2}+20\frac{Li(\lambda_{2}^3)}{\lambda_{2}^3}-
\left.20\frac{Li(\lambda_{2}^4)}{\lambda_{2}^4}\right)-
0.244\left(\frac{\alpha_{s}(Q^2)}{\alpha_{s}(\mu_{0}^2)}\right)^{364/405}\left(8\frac{Li(\lambda_2)}{\lambda_2}-\right.
112\frac{Li(\lambda_{2}^2)}{\lambda_{2}^2}+448\frac{Li(\lambda_{2}^3)}{\lambda_{2}^3}-
$$
\begin{equation}
672\frac{Li(\lambda_{2}^4)}{\lambda_{2}^4}+ \left.\left.
336\frac{Li(\lambda_{2}^5)}{\lambda_{2}^5}\right)\right],
\end{equation}
where $Li(\lambda)$ is the logarithmic integral for $\lambda>1$
defined as the principal value [55]
\begin{equation}
Li(\lambda)=P.V.\int_{0}^{\lambda}\frac{dx}{lnx},\,\,\,
\lambda_1=\hat s/\Lambda^2,  \lambda_2=-\hat u/\Lambda^2.
\end{equation}

 Hence, the effective cross section obtained after substitution of the
expressions (3.10-3.13) into the expression (2.15) is referred as
the running coupling effective cross section. We will denote the
higher-twist cross section obtained using the running coupling
constant approach by $(\Sigma_{M}^{HT})^{res}$.
\section{CONTRIBUTION OF THE  LEADING-TWIST DIAGRAMS}\label{lt}
Regarding the higher-twist corrections to the meson production cross
section, a comparison of our results with leading-twist
contributions is crucial. The contribution from the leading-twist
subprocess $\gamma\gamma\to q\overline{q}$ is shown in Fig.1(b). The
corresponding  inclusive cross section for production of a meson $M$
is given by
\begin{equation}
\left[\frac{d\sigma}{d^{3}p
}\right]_{\gamma\gamma \to
MX}=\frac{3}{\pi}\sum_{q,\overline{q}}\int_{0}^{1}\frac{dz}{z^2}
\delta(\hat{s}+\hat{t}+\hat{u})\hat{s}D_{q}^{M}(z,-\hat{t})\frac{d\sigma}
{d\hat{t}}(\gamma\gamma\to q\overline{q})
\end{equation}
where
$$
 \hat{s}=s,\,\,\hat{t}=\frac{t}{z}\,\,\,\hat{u}=\frac{u}{z}
$$
Here $s$, $t$, and $u$ refer to the overall $\gamma\gamma\to MX$
reaction. $D_{q}^{M}(z,-\hat t)$ represents the quark fragmentation
function into a meson containing a quark of the same flavor. For
$\pi^{+}$ production we assume
$D_{\pi^{+}/u}=D_{\pi^{+}/\overline{d}}$. In the leading-twist
subprocess, meson is indirectly emitted from the quark with
fractional momentum $z$. The $\delta$ function may be expressed in
terms of the parton kinematic variables, and the $z$ integration may
then be done. The final form for the leading-twist contribution to
the large-$p_{T}$ meson production cross section in the process
$\gamma\gamma\to MX$ is
$$
\Sigma_{M}^{LT}\equiv
E\frac{d\sigma}{d^{3}P}=\frac{3}{\pi}\sum_{q,\overline{q}}\int_{0}^{1}\frac{dz}{z^2}\delta
(\hat{s}+\hat{t}+\hat{u})\hat{s}D_{q}^{M}(z,-\hat{t})\frac{d\sigma}
{d\hat{t}}(\gamma\gamma\to q\overline{q})=
$$
\begin{equation}
\frac{3}{\pi}\sum_{q,\overline{q}}\int_{0}^{1}d\frac{1}{z}\delta(s+\frac{1}{z}
(t+u))\hat{s}D_{q}^{M}(z,-\hat{t})\frac{d\sigma}{d\hat{t}}(\gamma\gamma\to
q\overline{q})=
\frac{34}{27}\alpha_{E}^2\frac{1}{z}D_{q}^{M}(z)\frac{1}{{\hat
s}^2}\left[\frac{\hat t}{\hat u}+\frac{\hat u}{\hat t}\right]
\end{equation}
where
$$
 z=-\frac{t+u}{s}
$$
We should note that $D(z,-\hat t)/z$ behaves as $1/z^2$ as
$z\rightarrow0$. For the kinematic range considered in our numerical
calculations, $D(z,-\hat t)/z$ increases even more rapidly. We
obtain from the final cross section Eq.(4.2), following conclusion:
At fixed $p_T$, the cross section decreases with $s$ asymptotically
as $1/s$. At fixed $s$, the $D(z,-\hat t)$ function causes the cross
section to decrease rapidly as $p_T$ increases towards the
phase-spase boundary $(z\rightarrow1)$. As $s$ increases, the
phase-spase boundary  moves to higher $p_T$, and the $p_T$
distribution broadens.

\section{NUMERICAL RESULTS AND DISCUSSION}\label{results}

In this section, we discuss the numerical results for higher-twist
effects with higher-twist contributions calculated in the context
of the running coupling constant  and frozen coupling approaches
on the dependence of the chosen meson wave functions in the
process $\gamma \gamma \to MX$. We have calculated the dependence
on the meson wave functions for the higher-twist contribution to
the  large-$p_T$ single pseudoscalar ${\pi^{+}}$ and vector
$\rho_{L}^{+}$, $\rho_{T}^{+}$ mesons production cross section in
the photon-photon collision. The ${\pi^{-}}$, ${\rho_{L}^{-}}$,
${\rho_{T}^{-}}$ cross sections are, of course, identical. In the
calculations, we use the asymptotic wave function $\Phi_{asy}$,
the Chernyak-Zhitnitsky $\Phi_{CZ}$, the pion wave function from
which two nontrivial Gegenbauer coefficients $a_2$ and $a_4$ have
been extracted from the CLEO data on the $\pi^{0}\gamma$
transition form factor[46], the Braun-Filyanov pion wave functions
[11], and the Bakulev-Mikhailov-Stefanis pion wave function[ 47].
For $\rho$-meson we used Ball-Braun wave function[51]. For the
higher-twist subprocess, we take $\gamma
q_{1}\to(q_1\overline{q}_2)q_2$,
$\gamma\overline{q}_{2}\to(q_1\overline{q}_2)\overline{q}_2$
contributing to $\gamma\gamma\to MX$ cross sections. Inclusive
meson photoproduction represents a significant test case in which
higher-twist terms dominate those of leading-twist in certain
kinematic domains. For the dominant leading-twist subprocess for
the meson production, we take the photon-photon annihilation
$\gamma\gamma \to q\bar{q}$, in which the $M$ meson is indirectly
emitted from the quark. For example, the quark distribution
function inside the photon has been used [56]. The higher-twist
subprocesses probe the meson wave functions over a large range of
$Q^2$ squared momentum transfer, carried by the gluon. Therefore,
in the diagram given in Fig.1a we take $Q_{1}^2=x_{2}{\hat s}$,
$Q_{2}^2=-x_{1}\hat u$ , which we have obtained directly from the
higher-twist subprocesses diagrams. The same $Q_{1,2}^2$ has been
used as an argument of $\alpha_s(Q_{1,2}^2)$ in the calculation of
diagram.

The results of our numerical calculations are plotted in
Figs.2-33. First of all, it is very interesting to compare the
resummed higher-twist cross sections with the ones obtained in the
framework of the frozen coupling approach. In Figs.2-4 we show the
dependence of higher-twist cross sections
$(\Sigma_{M^{+}}^{HT})^{0}$ calculated in the context of the
frozen coupling, $(\Sigma_{M^{+}}^{HT})^{res}$ in the context of
the running coupling constant approaches and also the ratio
$R=(\Sigma_{M^{+}}^{HT})^{res}$/$\Sigma_{M^{+}}^{HT})^{0}$ as a
function of the meson transverse momentum $p_{T}$ for different
meson wave functions at $y=0$. It is seen that the values of cross
sections $(\Sigma_{M^{+}}^{HT})^{0}$,
$(\Sigma_{M^{+}}^{HT})^{res}$, and $R$ for fixed $y$ and $\sqrt s$
depend on the choice of the meson wave function. As seen from
Figs.2-3 in both cases, frozen coupling and running coupling
constant approaches the higher-twist differential cross section is
monotically decreasing with an increase in the transverse momentum
of the meson.  As is seen from Fig.4, when the transverse momentum
of the meson is increasing, the ratio $R$  is decreasing. But, as
shown in Fig.4, in the region $5\,\,GeV/c<p_T<80\,\,GeV/c$
higher-twist cross section calculated in the context of the
running coupling method is suppressed by about 2-4 orders of
magnitude relative to the higher-twist cross section calculated in
the framework of the frozen coupling method.
 In Figs.5 and 6, we shows the dependence of the ratio
$(\Sigma_{M^{+}}^{HT})^{0}$/$(\Sigma_{M^{+}}^{LT})$ and
$(\Sigma_{M^{+}}^{HT})^{res}$/$(\Sigma_{M^{+}}^{LT})$ as a
function of the meson transverse momentum $p_{T}$ for different
meson wave functions. Here  $(\Sigma_{M^{+}}^{LT})$ is the
leading-twist cross section, respectively. As seen from Fig.6, in
the region $5\,\,GeV/c<p_T<10\,\,GeV/c$  higher-twist cross
section for $\Phi_{BMS}(x,Q^2))$ and $\Phi_{CLEO}(x,Q^2))$  wave
functions calculated in the context of the running coupling method
is suppressed by about one orders of magnitude relative to the
leading-twist cross section, but in the region
$10\,\,GeV/c<p_T\leq 90\,\,GeV/c$ ratio is decreasing with  an
increase in the transverse momentum of  the meson. In Figs.7-9 we
show the dependence $(\Delta_{M}^{HT})^{0}$,
$(\Delta_{M}^{HT})^{res}$, and the ratio
$r$=($\Delta_{M}^{HT})^{res}$/$(\Delta_{M}^{HT})^{0}$, as a
function of the meson transverse momentum $p_{T}$ for the
different meson wave functions. Here,
$(\Delta_{M}^{HT})^0=(\Sigma_{M^{+}}^{HT})^0-
(\Sigma_{M^{-}}^{HT})^0$ and
$(\Delta_{M}^{HT})^{res}=(\Sigma_{M^{+}}^{HT})^{res}-
(\Sigma_{M^{-}}^{HT})^{res}$. As seen from Figs.7 and 8, the
difference of the higher-twist differential cross section is
decreasing with an increase in the transverse momentum of the
meson. The dependence, as shown in Fig.9, is identically
equivalent to Fig.4. In Figs.10-17, we have depicted higher-twist
cross sections, ratios $(\Sigma_{M^{+}}^{HT})^{0}$,
$(\Sigma_{M^{+}}^{HT})^{res}$,
$R=(\Sigma_{M^{+}}^{HT})^{res}$/$(\Sigma_{M^{+}}^{HT})^{0}$,
$r$=($\Delta_{M}^{HT})^{res}$/$(\Delta_{M}^{HT})^{0}$,
$(\Delta_{M}^{HT})^{0}$, $(\Delta_{M}^{HT})^{res}$,
$(\Sigma_{M^{+}}^{HT})^{0}$/$(\Sigma_{M^{+}}^{LT})$ and
$(\Sigma_{M^{+}}^{HT})^{res}$/$(\Sigma_{M^{+}}^{LT})$  as a
function of the rapidity $y$ of the meson at $\sqrt s=183\,\,GeV$
and $p_T=14.6\,\,GeV/c$. At $\sqrt s=183\,\,GeV$ and
$p_T=14.6\,\,GeV/c$, the meson rapidity lies in the region
$-2.52\leq y\leq2.52$.

As seen from Fig.10 and Fig.14, in the region ($-2.52\leq y\leq
-1.92$), the cross section for all wave functions increases with an
increase of the $y$ rapidity of the meson and have a maximum
approximately at the point $y=-1.92$. Besides that, the cross
sections decrease with an increase in the $y$ rapidity of the meson.
But, as seen from Figs.12 and 13 in the region ($-2.52\leq y\leq
1.92$) the cross section for all wave functions increases with an
increase of the $y$ rapidity of the meson and has a maximum
approximately at the point $y=1.92$. But, as seen from Figs.16-17 in
the region ($-2.52\leq y\leq 1.92$) the cross section for all wave
functions has a minimum approximately at the point $y=1.92$. As is
seen from Figs.10-17, cross sections, the ratios $R$ and $r$ are
very sensitive to the choice of the meson wave functions.  It should
be noted that the magnitude of the higher-twist cross section for
the pion wave functions $\Phi_{BMS}(x,Q^2)$ and $\Phi_{CLEO}(x,Q^2)$
is very close to the asymptotic wave function $\Phi_{asy}(x)$. Also,
the distinction between $R(\Phi_{asy}(x))$ with
$R(\Phi_{BMS}(x,Q^2))$, $R(\Phi_{CZ}(x,Q^2))$,
$R(\Phi_{CLEO}(x,Q^2))$, $R(\Phi_{BF}(x,Q^2))$,
$R(\Phi_{BB(L)}(x,Q^2))$  and $R(\Phi_{BB(T)}(x,Q^2))$ have been
calculated. For example, in the case of $\sqrt s=183\,\,GeV$, $y=0$,
the distinction between $R(\Phi_{asy}(x))$ with
$R(\Phi_{i}(x,Q^2))$\,\,[i=BMS, CZ, CLEO, BF, BB(L), BB(T)] as a
function of the meson transverse momentum $p_{T}$ is shown in Table
\ref{table1}. Thus, the distinction between $R(\Phi_{asy}(x))$ with
$R(\Phi_{i}(x,Q^2))$,\,\,[i=BMS, CLEO] is maximum at
$p_T=5\,\,GeV/c$, with $R(\Phi_{CZ}(x))$  at  $p_T=50\,\,GeV/c$; the
distinction between $R(\Phi_{asy}(x))$ with $R(\Phi_{BF}(x,Q^2))$,
is maximum at $p_T=90\,\,GeV/c$; but the distinction
$R(\Phi_{asy}(x))$ with $R(\Phi_{i}(x,Q^2))$,\,\,[i=BB(L), BB(T)] is
maximum at $p_T=75\,\,GeV/c$, Also, we have calculated the
distinction between $R(\Phi_{asy}(x))$ with
$R(\Phi_{i}(x,Q^2))$\,\,[i=BMS, CZ, CLEO, BF, BB(L), BB(T)] as a
function of the rapidity $y$ of the meson. For example, in the case
of $\sqrt s=183GeV$, $p_{T}=14.6GeV/c$  the distinction between
$R(\Phi_{asy}(x))$ with $R(\Phi_{i}(x,Q^2))$ \,\,[i=BMS, CZ, CLEO,
BF, BB(L), BB(T)] as a function of the rapidity $y$ of the meson is
presented in Table \ref{table2}

We have also carried out comparative calculations in the
center-of-mass energy $\sqrt s=209\,\,GeV$. The results of our
numerical calculations in the center-of-mass energies $\sqrt
s=209\,\,GeV$ are plotted in Figs.18-33. Analysis of our
calculations at the center-of-mass energies $\sqrt s=183\,\,GeV$
and $\sqrt s=209\,\,GeV$, show that with the increase in beam
energy values of the cross sections, ratio
$R=(\Sigma_{M^{+}}^{HT})^{res}/(\Sigma_{M^{+}}^{HT})^{0}$, and
contributions of higher-twist to the cross section decrease by
about 1-2 order. Therefore the experimental investigation  of
higher-twist effects include renormalon effects conveniently in
low energy. On the other hand, the higher-twist corrections and
ratios $R$ and $r$ are very sensitive to the choice of the meson
wave function. Also, the distinction between $R(\Phi_{asy}(x))$
with $R(\Phi_{BMS}(x,Q^2))$, $R(\Phi_{CZ}(x,Q^2))$,
$R(\Phi_{CLEO}(x,Q^2))$, $R(\Phi_{BF}(x,Q^2))$,
$R(\Phi_{BB(L)}(x,Q^2))$  and $R(\Phi_{BB(T)}(x,Q^2))$ have been
calculated. For example, in the case of $\sqrt s=209\,\,GeV$,
$y=0$, the distinction between $R(\Phi_{asy}(x))$ with
$R(\Phi_{i}(x,Q^2))$\,\,[i=BMS, CZ, CLEO, BF, BB(L), BB(T)] as a
function of the meson transverse momentum $p_{T}$ is shown in
Table \ref{table3}. Thus, the distinction between
$R(\Phi_{asy}(x))$ with $R(\Phi_{i}(x,Q^2))$,\,\,(i=BMS, CLEO) is
maximum at $p_T=10\,\,GeV/c$, with $R(\Phi_{CZ}(x))$  at
$p_T=65\,\,GeV/c$; the distinction between $R(\Phi_{asy}(x))$ with
$R(\Phi_{BF}(x,Q^2))$, is maximum at $p_T=100\,\,GeV/c$; but the
distinction $R(\Phi_{asy}(x))$ with
$R(\Phi_{i}(x,Q^2))$,\,\,(i=BB(L), BB(T)) is maximum at
$p_T=65\,\,GeV/c$. Also, we have calculated the distinction
between $R(\Phi_{asy}(x))$ with $R(\Phi_{i}(x,Q^2))$\,\,[i=BMS,
CZ, CLEO, BF, BB(L), BB(T)] as a function of the rapidity $y$ of
the meson. For example, in the case of $\sqrt s=209GeV$,
$p_{T}=16.7GeV/c$  the distinction between $R(\Phi_{asy}(x))$ with
$R(\Phi_{i}(x,Q^2))$ \,\,[i=BMS, CZ, CLEO, BF, BB(L), BB(T)] as a
function of the rapidity $y$ of the meson is presented in Table
\ref{table4}. The calculations show that the ratio
$R(\Phi_{i}(x,Q^2))$/$R(\Phi_{asy}(x))$, (i=CLEO, CZ, BF, BMS,
BB(L), BB(T)) for all values of the transverse momentum $p_T$ of
the meson identically equivalent to ratio
$r(\Phi_{i}(x,Q^2))$/$r(\Phi_{asy}(x))$. Results of our numerical
calculations demonstrate that in the renormalon approach there are
not difference between results obtained using the cross sections
(2.13), (2.14) or (2.11), (2.12).

The total integrated luminosity of LEP is 612.8$pb^{-1}$ and total
luminosity of ILC required is 500$fb^{-1}$, also a peak luminosity
of ILC is 1000 $fb^{-1}$ during the first phase of operation at
$209GeV\div500GeV$. In our calculations of the higher-twist cross
section of the process the dependence of the transverse momentum of
meson appears in the range of $(10^{-10}\div10^{-24})mb/GeV^2$, or
$(10^{-1}\div10^{-15})pb/GeV^2$. Therefore, higher-twist cross
section obtained in our paper should be observable at LEP and ILC.

\section{Concluding Remarks}\label{conc}
In this work we have calculated the single meson inclusive
production via higher-twist mechanism and obtained the expressions
for the subprocess $\gamma q \to Mq$ cross section for mesons with
symmetric wave functions. For calculation of the cross section we
have applied the running coupling constant method and revealed
infrared renormalon poles in the cross section expression. Infrared
renormalon induced divergences have been regularized by means of the
principal value prescription and the resummed expression (the Borel
sum) for the higher-twist cross section has been found. The
higher-twist cross sections were calculated in the frozen coupling
and running coupling approaches. The resummed higher-twist cross
section differs from that found using the frozen coupling approach,
in some regions, considerably. Also we demonstrated that
higher-twist contributions to single meson production cross section
in the photon-photon collisions have important phenomenological
consequences. We have obtained very interesting results. The ratio
$R$ for all values of the transverse momentum $p_{T}$ and of the
rapidity $y$ of the meson identically equivalent to ratio $r$. Our
investigation enables us to conclude that the higher-twist meson
production cross section in the photon-photon collisions depends on
the form of the meson model wave functions and may be used for their
study. Analysis of our calculations shows that the magnitude of
cross sections of the leading-twist is larger than the higher-twist
cross sections ones calculated in the frozen coupling approach in
2-4 order. But, in some regions of transverse momentum of the meson,
the higher-twist cross section calculated in the context of the
running coupling method is comparable with the cross sections of
leading-twist. Further investigations are needed in order to clarify
the role of high twist effects  in this process. We have
demonstrated that the resummed result depends on the meson model
wave functions used in calculations.  The production of high-$p_{T}$
meson probes the short-distance dynamics of photon-photon reactions.
In addition to providing tests of perturbative QCD, $\gamma \gamma$
processes with real or almost real photons give us information on
the photon structure function  which is complementary to the
information gained from deep inelastic scattering on a real photon.
The latter process essentially probes the quark distribution while
high-$p_{T}$ meson production is also sensitive to the gluon
distribution of the photon. As it is well known high-$p_{T}$
processes induced by real photons have a rather complex structure.
This arises from the fact that the photon couples to the hard
subprocess either directly or through its quark and gluon content.
In particular, meson production in photon-photon collisions  takes
into account infrared renormalon effects: this opens a window
toward new types of photon structure function which can not be
measured by the lepton-photon scatterings.


\section*{Acknowledgments}
One of author, A. I.~Ahmadov is grateful to Prof. Martin Beneke
and also other members of the Institut f\"{u}r Theoretische Physik
E, RWTH Aachen University  for appreciates hospitality extended to
him in Aachen, where this work has been carried out and to
Deutscher Akademischer Austausch Dienst (DAAD) for financial
support. A.I.Ahmadov also thanks Dr.S.S.Agaev for drawing to his
attention deficiences in the first version of this paper.

\newpage

\begin{table}[ht]
\begin{center}
\begin{tabular}{|c|c|c|c|c|c|c|c} \hline
$p_{T},GeV/c$ & $\frac{R(\Phi_{BMS}(x,Q^2))}{R(\Phi_{asy}(x))}$ &
$\frac{R(\Phi_{CZ}(x,Q^2))}{R(\Phi_{asy}(x))}$ &
$\frac{R(\Phi_{CLEO}(x,Q^2))}{R(\Phi_{asy}(x))}$ &
$\frac{R(\Phi_{BF}(x,Q^2))}{R(\Phi_{asy}(x))}$ &
$\frac{R(\Phi_{BB(L)}(x,Q^2))}{R(\Phi_{asy}(x))}$&
$\frac{R(\Phi_{BB(T)}(x,Q^2))}{R(\Phi_{asy}(x))}$ \\ \hline
  5 & 132.678  & 0.288 & 89.019  &7.621 & 0.499&0.0808& \\ \hline
  30 & 3.064  & 0.949 & 7.24 &29.651 & 1.103 &0.167& \\ \hline
  50& 2.46  & 2.516 & 7.955 & 30.635 & 1.962 &0.343&\\ \hline
  75 & 2.992  & 1.778 & 3.12  &92.936 & 2.091&0.359& \\ \hline
  90 & 13.888  & 0.293 & 6.744 &157.748 & 0.712 &0.116 \\ \hline
\end{tabular}
\end{center}
\caption{The distinction between $R(\Phi_{asy}(x))$ with
$R(\Phi_{i}(x,Q^{2}))$  (i=BMS, CZ, CLEO, BF, BB(L), BB(T) ) at c.m.
 energy $\sqrt s=183GeV$.} \label{table1}
\end{table}

\begin{table}[ht]
\begin{center}
\begin{tabular}{|c|c|c|c|c|c|c|c}\hline
$y$ & $\frac{R(\Phi_{BMS}(x,Q^2))}{R(\Phi_{asy}(x))}$
&$\frac{R(\Phi_{CZ}(x,Q^2))}{R(\Phi_{asy}(x))}$ &
$\frac{R(\Phi_{CLEO}(x,Q^2))}{R(\Phi_{asy}(x))}$ &
$\frac{R(\Phi_{BF}(x,Q^2))}{R(\Phi_{asy}(x))}$ &
$\frac{R(\Phi_{BB(L)}(x,Q^2))}{R(\Phi_{asy}(x))}$&
$\frac{R(\Phi_{BB(T)}(x,Q^2))}{R(\Phi_{asy}(x))}$\\ \hline
  -2.52 & 0.0307 & 0.259 & 7.091  &0.6706 & 0.751&0.048& \\ \hline
  -1.92 & 0.338  & 4.052 & 9.741 &4.327 & 1.965 &0.448& \\ \hline
  0.78& 14.858  & 0.344 & 34.922 & 13.642 & 0.586 &0.0794&\\ \hline
  1.38 & 18.125  & 0.309 & 40.788  & 15.9209 & 0.477&0.0792 \\ \hline
  2.28 & 0.9125  & 0.327 & 3.3298 & 1.0724 & 0.8047 &0.0895& \\ \hline
\end{tabular}
\end{center}
\caption{The distinction between $R(\Phi_{asy}(x))$ with
$R(\Phi_{i}(x,Q^{2}))$ (i=BMS, CZ, CLEO, BF, BB(L), BB(T)) at c.m.
energy $\sqrt s=183GeV$.} \label{table2}
\end{table}

\begin{table}[ht]
\begin{center}
\begin{tabular}{|c|c|c|c|c|c|c|c}\hline
$p_{T},GeV/c$ & $\frac{R(\Phi_{BMS}(x,Q^2))}{R(\Phi_{asy}(x))}$ &
$\frac{R(\Phi_{CZ}(x,Q^2))}{R(\Phi_{asy}(x))}$ &
$\frac{R(\Phi_{CLEO}(x,Q^2))}{R(\Phi_{asy}(x))}$ &
$\frac{R(\Phi_{BF}(x,Q^2))}{R(\Phi_{asy}(x))}$ &
$\frac{R(\Phi_{BB(L)}(x,Q^2))}{R(\Phi_{asy}(x))}$ &
$\frac{R(\Phi_{BB(T)}(x,Q^2))}{R(\Phi_{asy}(x))}$\\ \hline
  10 &33.874  & 0.352 & 33.404 &11.184 & 0.568&0.08812 \\ \hline
  35 & 2.537  & 0.9789 & 6.7409 &29.866 & 1.1213&0.1704&  \\ \hline
  65& 2.167  & 3.0311 & 7.7455 & 30.915 & 2.2177 &0.4091\\ \hline
  85 & 2.358  & 1.841 & 3.0529  &86.0547 & 2.0994&0.3638& \\ \hline
  100 & 6.5605 & 0.4398 & 3.0772 & 194.339 & 1.1648 &0.1833& \\ \hline
\end{tabular}
\end{center}
\caption{The distinction between $R(\Phi_{asy}(x))$ with
$R(\Phi_{i}(x,Q^{2}))$  (i=BMS, CZ, CLEO, BF, BB(L), BB(T)) at c.m.
energy $\sqrt s=209 GeV$.}\label{table3}
\end{table}

\newpage

\begin{table}[ht]
\begin{center}
\begin{tabular}{|c|c|c|c|c|c|c|c}\hline
$y$ & $\frac{R(\Phi_{BMS}(x,Q^2))}{R(\Phi_{asy}(x))}$ &
$\frac{R(\Phi_{CZ}(x,Q^2))}{R(\Phi_{asy}(x))}$  &
$\frac{R(\Phi_{CLEO}(x,Q^2))}{R(\Phi_{asy}(x))}$ &
$\frac{R(\Phi_{BF}(x,Q^2))}{R(\Phi_{asy}(x))}$ &
$\frac{R(\Phi_{BB(L)}(x,Q^2))}{R(\Phi_{asy}(x))}$&
$\frac{R(\Phi_{BB(T)}(x,Q^2))}{R(\Phi_{asy}(x))}$ \\ \hline
  -2.25 & 0.0223 & 0.2596 & 6.0611 &0.7024 & 0.7847&0.0639&\\ \hline
  -1.92 & 0.2747& 3.8857 & 8.6517& 4.122& 1.9435&0.2391&  \\ \hline
  0.78& 12.9913 & 0.3491 & 32.4146& 7.0045& 0.5867&0.07707&\\ \hline
  1.38 & 15.879  & 0.3138 & 37.443& 2.7794& 0.475&0.1251&\\ \hline
  2.28 & 0.7783 & 0.3323 & 2.9838 & 0.4545 & 0.8104&0.2131  \\ \hline
\end{tabular}
\end{center}
\caption{The distinction between $R(\Phi_{asy}(x))$ with
$R(\Phi_{i}(x,Q^{2}))$ (i=BMS, CZ, CLEO, BF, BB(L), BB(T)) at c.m.
energy $\sqrt s=209GeV$.} \label{table4}
\end{table}

\begin{figure}[hpb]
\includegraphics[width=8cm]{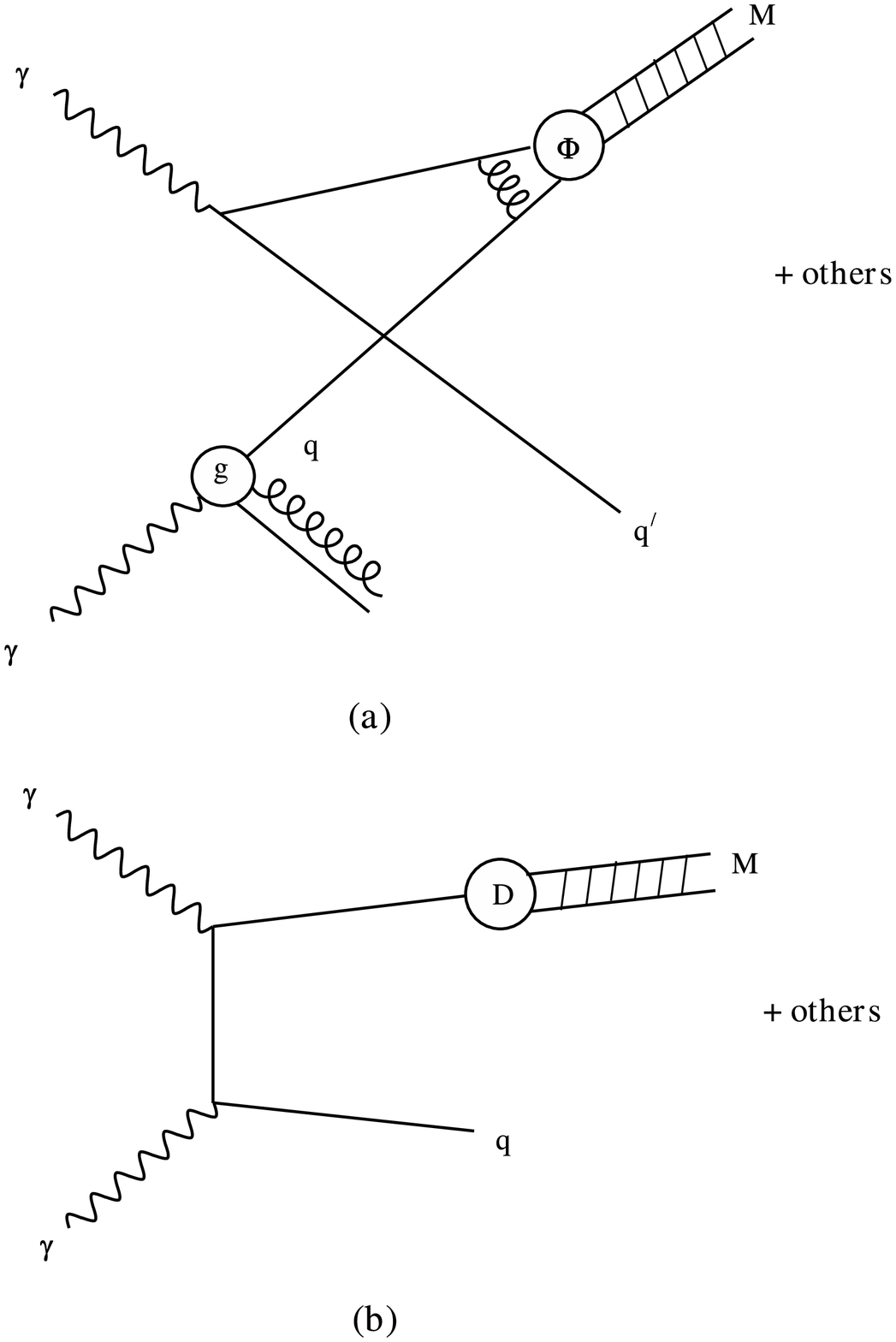} \vskip -0.02cm
\caption{(a): The higher-twist contribution to $\gamma\gamma\to
MX$;\,\,\,(b): The leading-twist contribution to $\gamma\gamma\to
MX$} \label{Fig1}
\end{figure}

\newpage

\begin{figure}[htb]
\includegraphics[width=12.8cm]{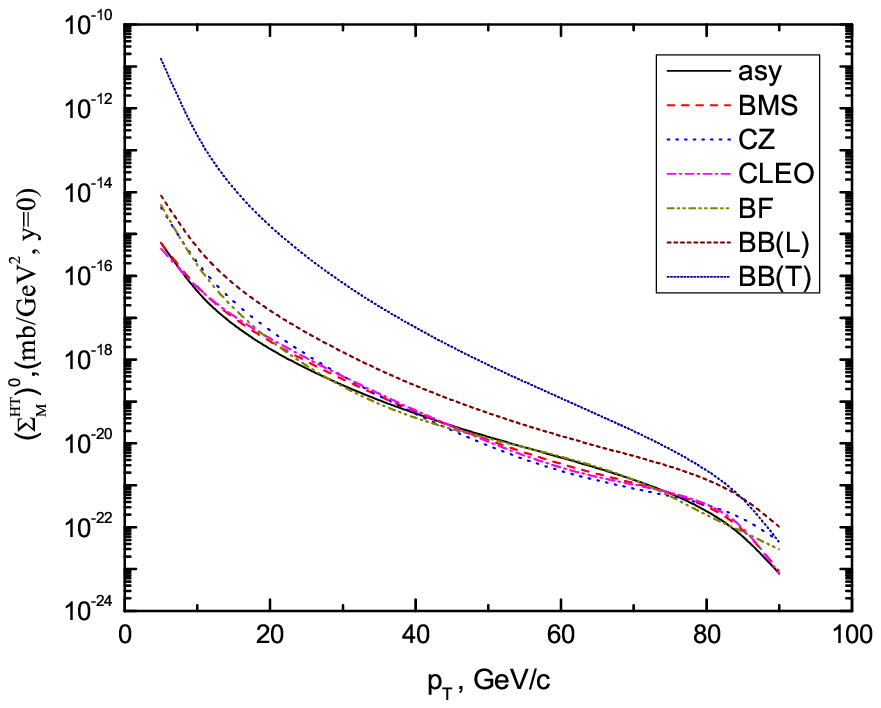}
\vskip -1.08cm \caption{Higher-twist $M$ production cross section
$(\Sigma^{HT})^{0}$ as a function of the $p_{T}$ transverse
momentum of the meson at the c.m.energy  $\sqrt s=183\,\,
GeV$.}\label{Fig2}
 \vskip-0.2cm
\includegraphics[width=12.8cm]{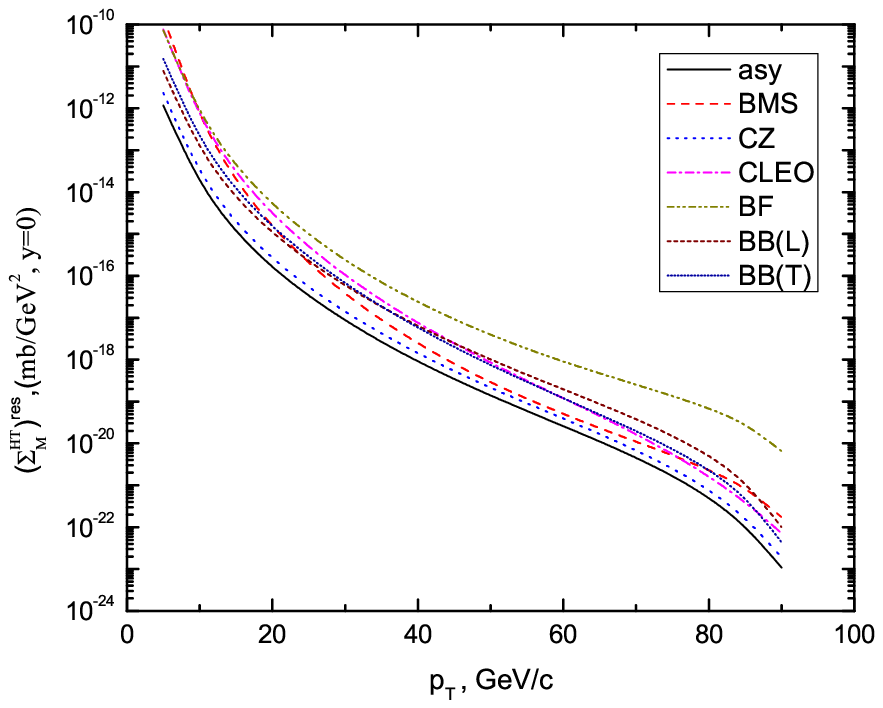}
\vskip -0.5cm \caption{Higher-twist $M$ production cross section
$(\Sigma^{HT})^{res}$ as a function of the $p_{T}$ transverse
momentum of the meson at the c.m.energy  $\sqrt s=183\,\, GeV$.}
\label{Fig3}
\end{figure}

\newpage

\begin{figure}[htb]
\includegraphics[width=12.8cm]{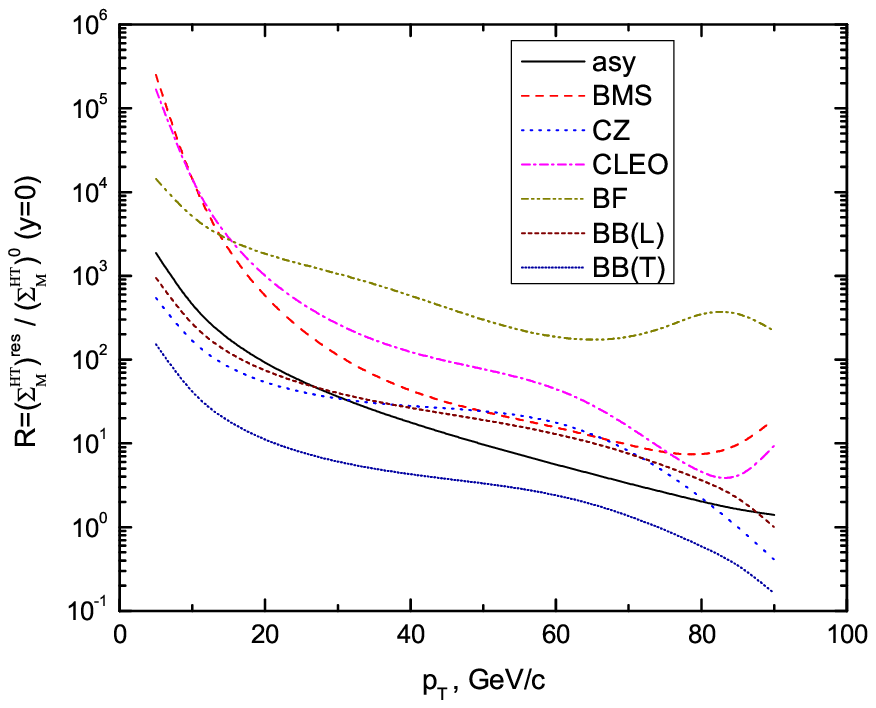}
 \vskip-1.08cm
\caption{Ratio
$R=(\Sigma_{M}^{HT})^{res}/(\Sigma_{M}^{HT})^{0}$, where
higher-twist contribution are calculated for the meson rapidity $y=0$
at the c.m.energy $\sqrt s=183\,\,GeV$ as a function of the meson
transverse momentum, $p_{T}$.} \label{Fig4}
 \vskip -0.2cm
\includegraphics[width=12.8cm]{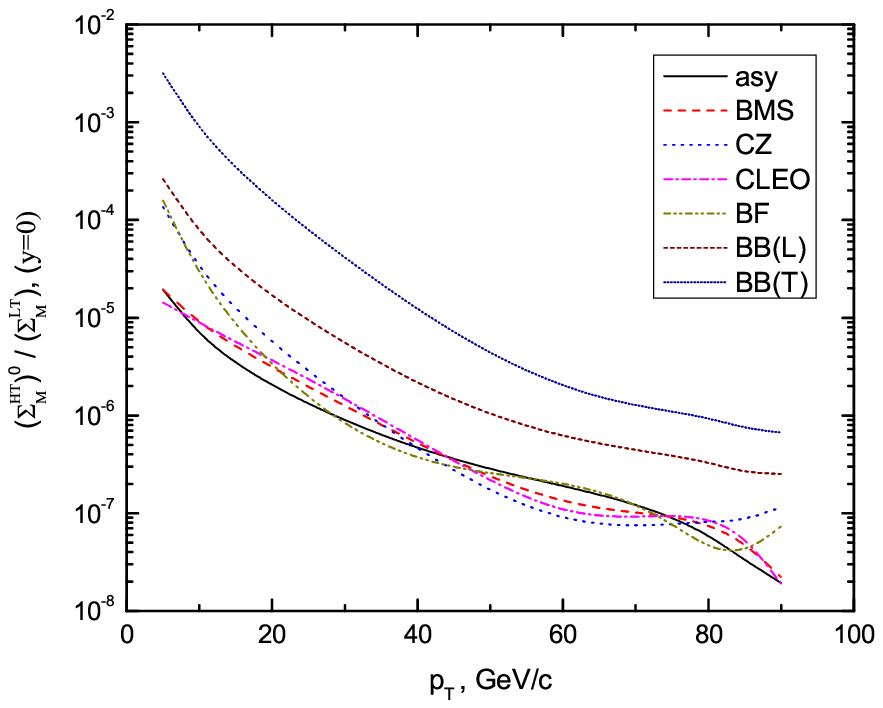}
  \vskip-0.5cm
\caption{Ratio $(\Sigma_{M}^{HT})^{0}/(\Sigma_{M}^{LT})$, as a
function of the  $p_{T}$ transverse momentum of the meson  at the
c.m. energy $\sqrt s=183\,\,GeV$.} \label{Fig5}
\end{figure}

\newpage

\begin{figure}[htb]
\includegraphics[width=12.8cm]{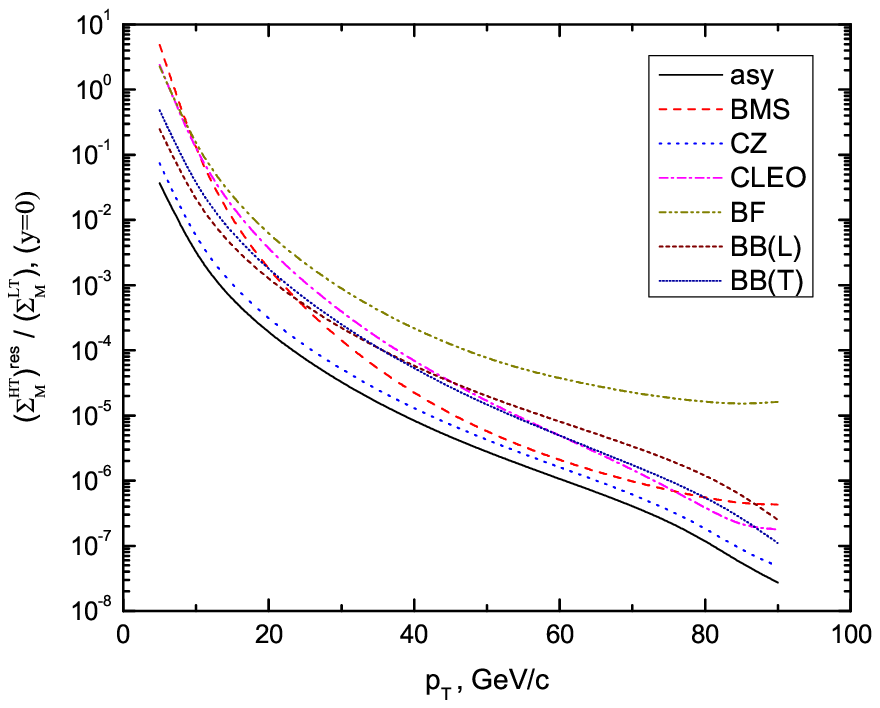}
\vskip-1.08cm \caption{Ratio
$(\Sigma_{M}^{HT})^{res}/(\Sigma_{M}^{LT})$, as a function of the
$p_{T}$ transverse momentum of the meson  at the c.m. energy
$\sqrt {s} =183\,\,GeV$.} \label{Fig6} \vskip-0.2cm
\includegraphics[width=12.8cm]{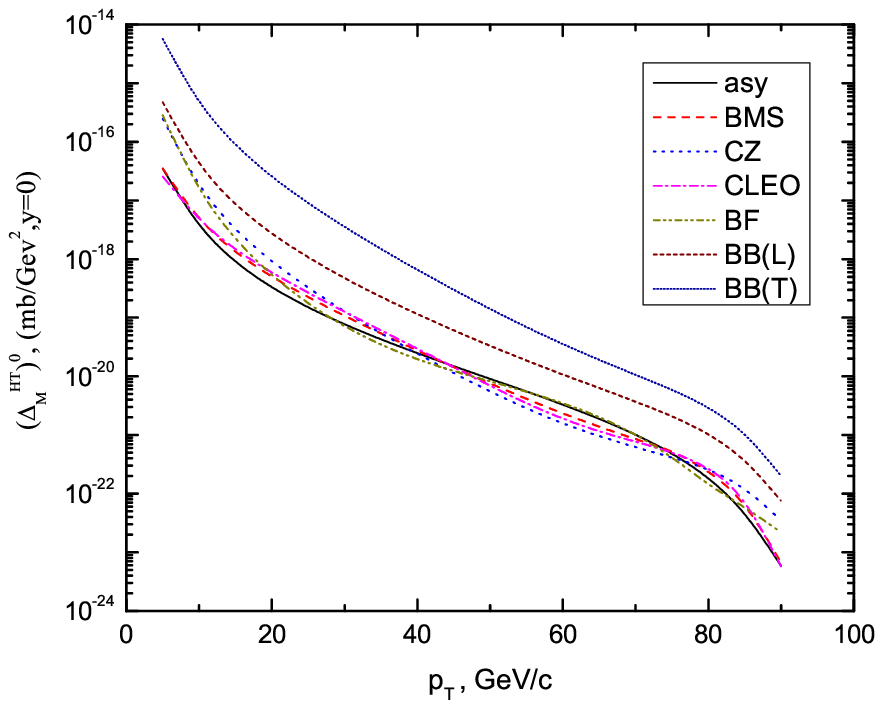}
\vskip-0.5cm \caption{The difference of the higher-twist cross
section,
$(\Delta_{M}^{HT})^{0}=(\Sigma_{M^{+}}^{HT})^{0}-(\Sigma_{M^{-}}^{HT})^{0}$,
as a function of the meson transverse momentum, $p_{T}$, at the
c.m.energy $\sqrt s=183\,\, GeV$.}\label{Fig7}
\end{figure}

\newpage

\begin{figure}[htb]
\includegraphics[width=12.8cm]{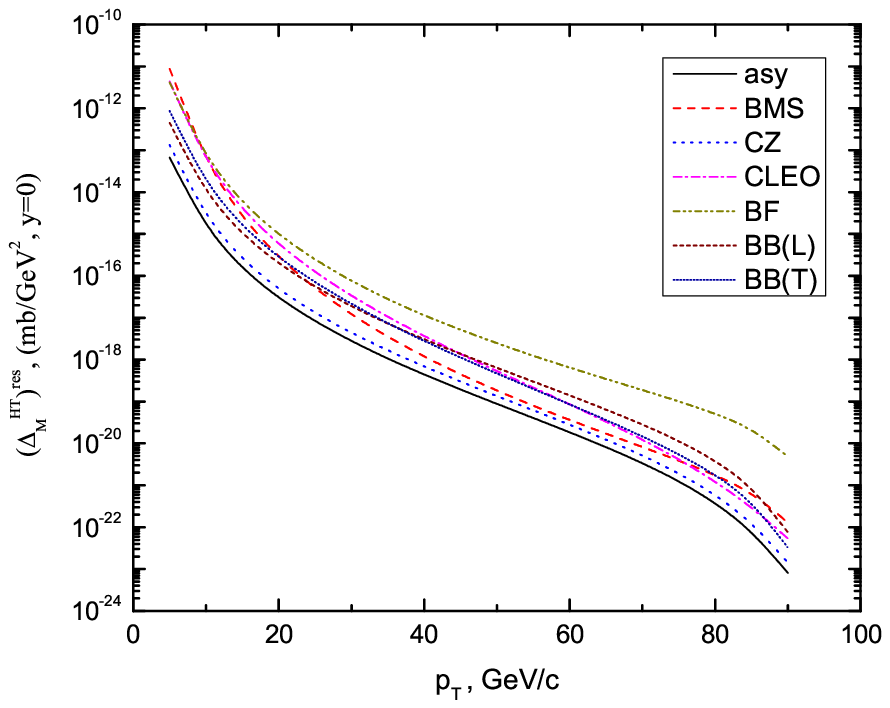}
\vskip-1.08cm \caption{The difference of the higher-twist cross
section,
$(\Delta_{M}^{HT})^{res}=(\Sigma_{M^{+}}^{HT})^{res}-(\Sigma_{M^{-}}^{HT})^{res}$,
as a function of the meson transverse momentum, $p_{T}$, at the
c.m.energy $\sqrt s=183\,\, GeV$.} \label{Fig8}
\vskip-0.2cm
\includegraphics[width=12.8cm]{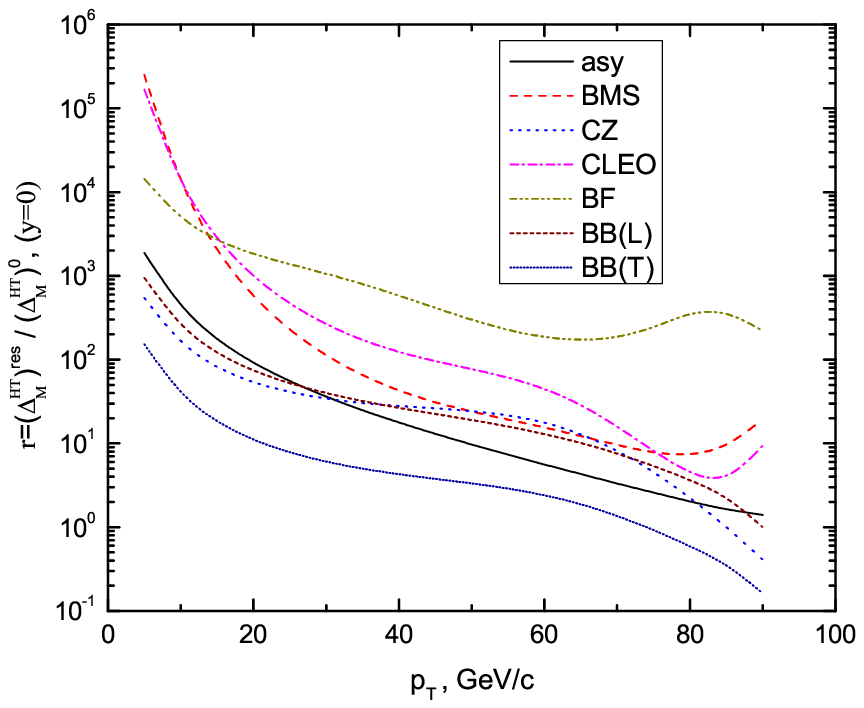}
\vskip-0.5cm \caption{Ratio
$r=(\Delta_{M}^{HT})^{res}/(\Delta_{M}^{HT})^{0}$, where higher-twist
contributions are calculated for the meson rapidity $y=0$ at
the c.m. energy $\sqrt s=183\,\, GeV$, as a function of the meson
transverse momentum, $p_T$.} \label{Fig9}
\end{figure}

\newpage

\begin{figure}[htb]
\includegraphics[width=12.8cm]{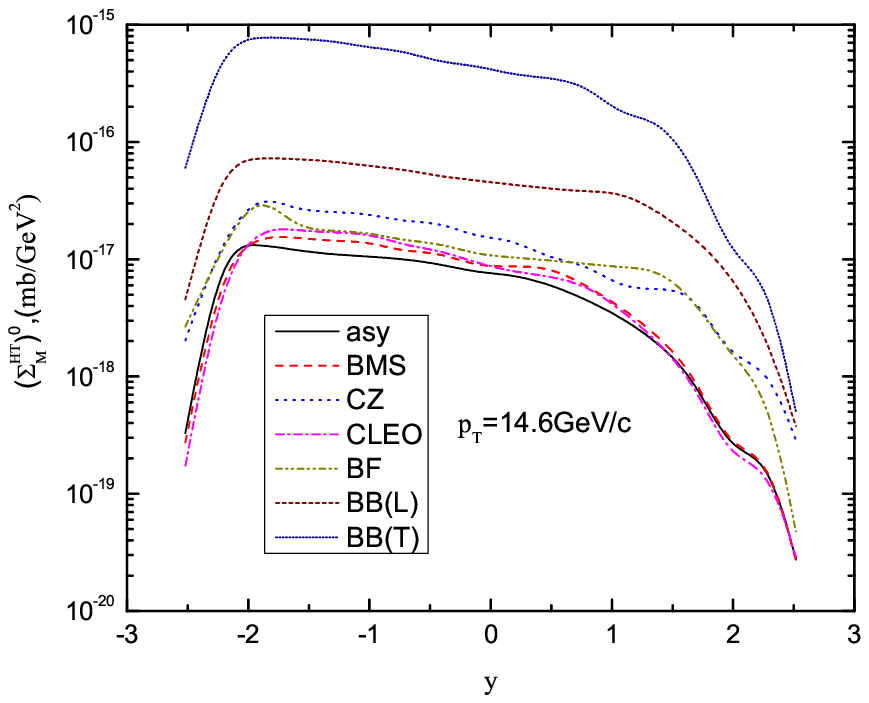}
\vskip-1.12cm \caption{Higher-twist $M$ production cross section
$(\Sigma_{M}^{HT})^{0}$, as a function of the $y$ rapidity of the
meson at the  transverse momentum of the meson $p_T=14.6\,\,
GeV/c$, at the c.m. energy $\sqrt s=183\,\, GeV$.} \label{Fig10}
\vskip-0.4cm
\includegraphics[width=12.8cm]{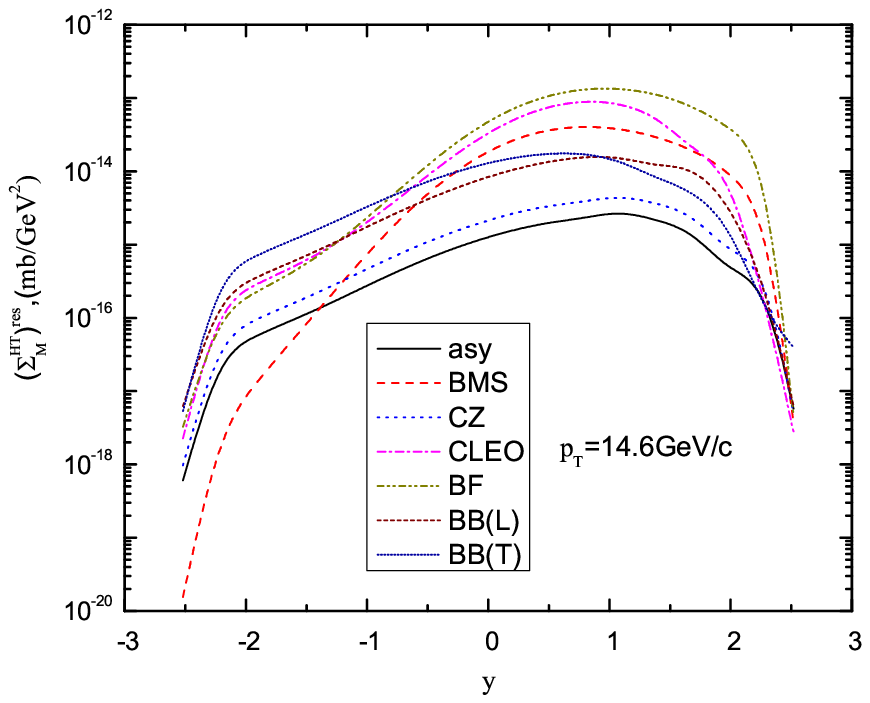}
 \vskip-0.8cm
\caption{Higher-twist $M$ production cross section
$(\Sigma_{M}^{HT})^{res}$, as a function of the $y$ rapidity of the
meson at the  transverse momentum of the meson $p_T=14.6\,\, GeV/c$,
at the c.m. energy $\sqrt s=183\,\, GeV$.} \label{Fig11}
\end{figure}

\newpage

\begin{figure}[htb]
\includegraphics[width=12.8cm]{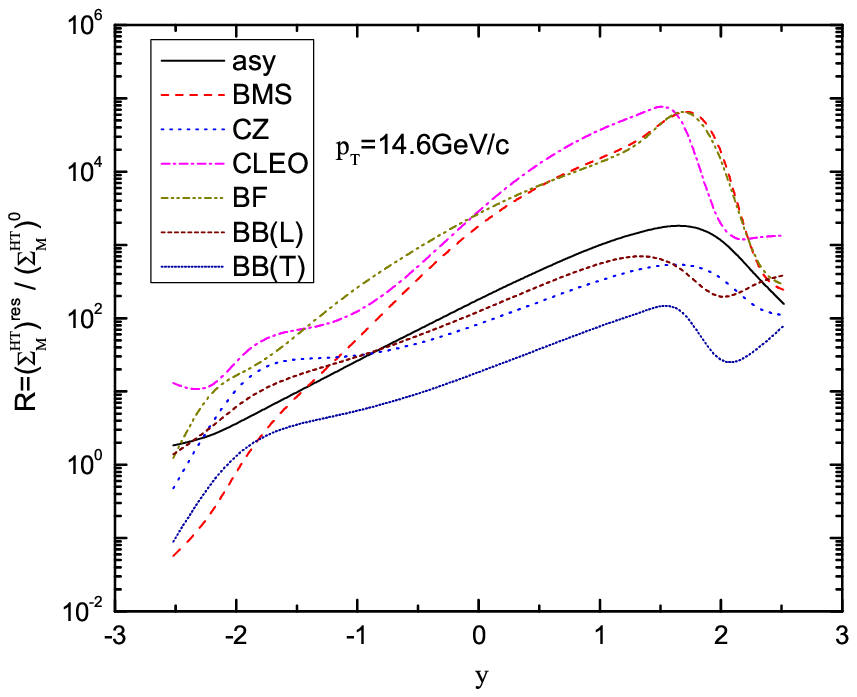}
\vskip-1.08cm \caption{Ratio
$R=(\Sigma_{M}^{HT})^{res}/(\Sigma_{M}^{HT})^{0}$, as a function
of the $y$ rapidity of the meson at the  transverse momentum of
the meson $p_T=14.6\,\, GeV/c$, at the c.m. energy $\sqrt
s=183\,\, GeV$.} \label{Fig12}\vskip-0.2cm
\includegraphics[width=12.8cm]{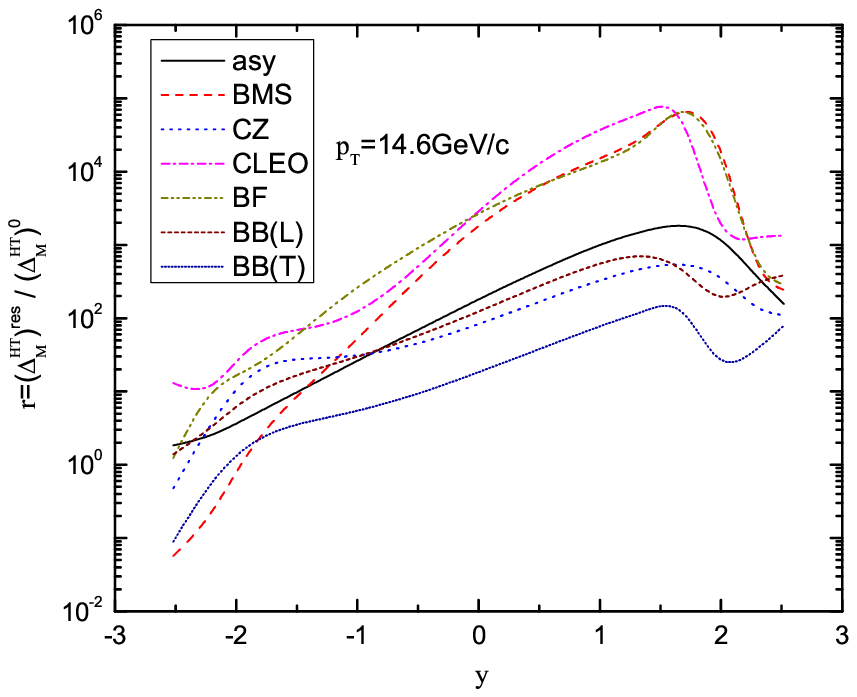}
\vskip-0.5cm \caption{Ratio
$r=(\Delta_{M}^{HT})^{res}/(\Delta_{M}^{HT})^{0}$, as a function
of the $y$ rapidity of the meson at the  transverse momentum of
the meson $p_T=14.6\,\, GeV/c$, at the c.m. energy $\sqrt
s=183\,\, GeV$} \label{Fig13}
\end{figure}

\newpage

\begin{figure}[htb]
\includegraphics[width=12.8cm]{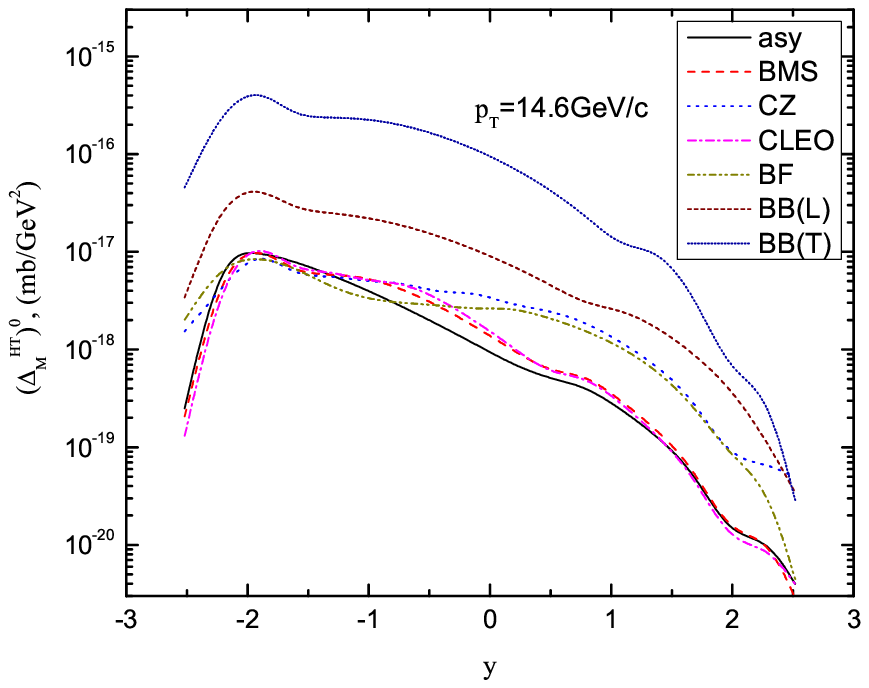}
\vskip-1.12cm \caption{The difference of the higher-twist cross
section,
$(\Delta_{M}^{HT})^{0}=(\Sigma_{M^{+}}^{HT})^{0}-(\Sigma_{M^{-}}^{HT})^{0}$,
as a function of the $y$ rapidity of the meson at the  transverse
momentum of the meson $p_T=14.6\,\, GeV/c$, at the c.m. energy
$\sqrt s=183\,\, GeV$.} \label{Fig14} \vskip -0.4cm
\includegraphics[width=12.8cm]{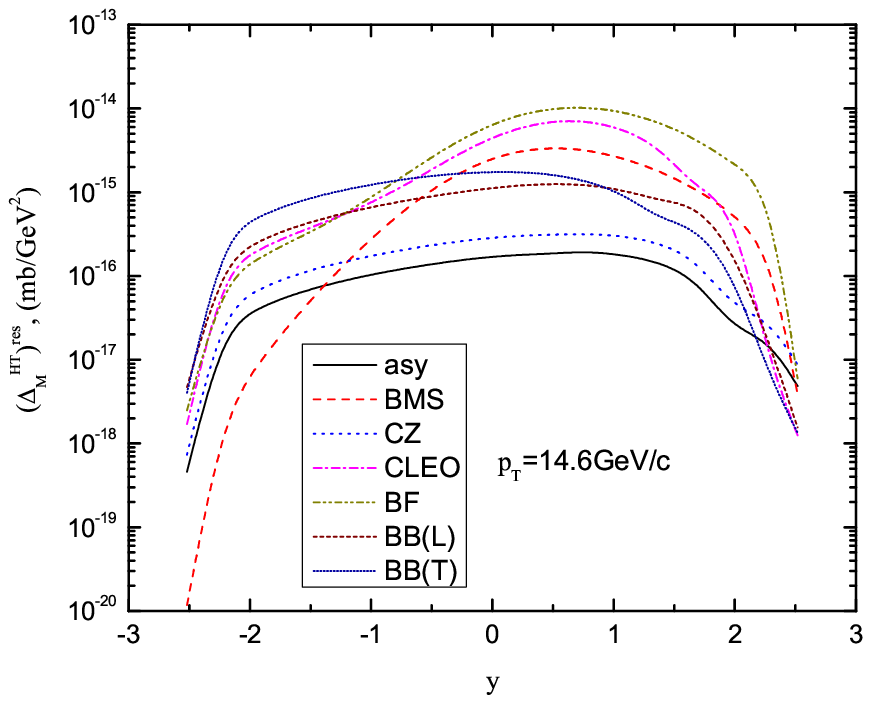}
 \vskip-0.8cm
\caption{The difference of the higher-twist cross
section,
$(\Delta_{M}^{HT})^{res}=(\Sigma_{M^{+}}^{HT})^{res}-(\Sigma_{M^{-}}^{HT})^{res}$, as a function of the
$y$ rapidity of the meson at the  transverse momentum of the meson
$p_T=14.6\,\, GeV/c$, at the c.m. energy $\sqrt s=183\,\, GeV$.}
\label{Fig15}
\end{figure}

\newpage
\clearpage

\begin{figure}[htb]
\includegraphics[width=12.8cm]{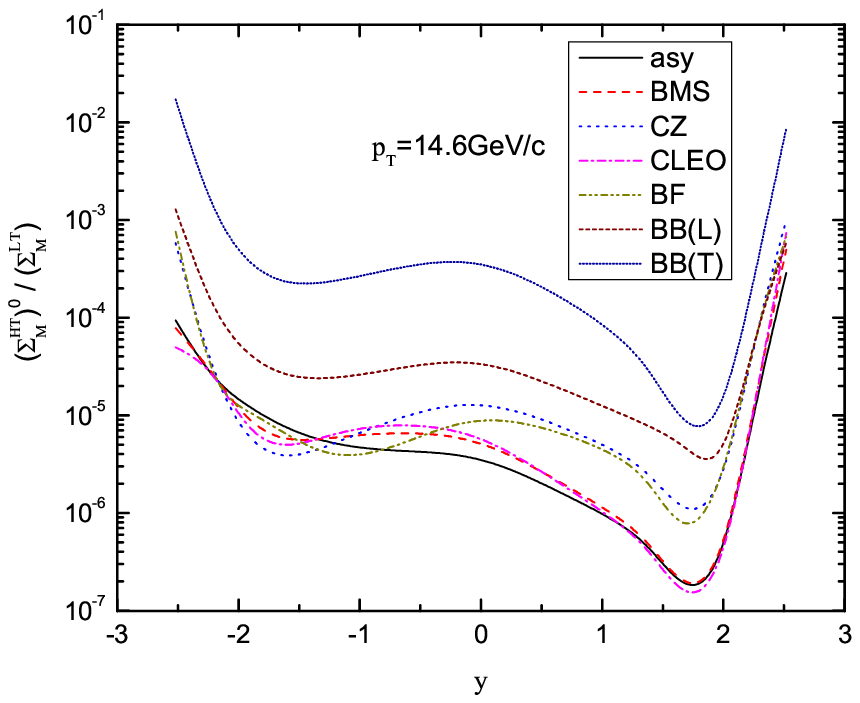}
\vskip-1.08cm \caption{Ratio
$(\Sigma_{M}^{HT})^{0}/(\Sigma_{M}^{LT})$, as a function of the
$y$ rapidity of the meson at the  transverse momentum of the meson
$p_T=14.6\,\, GeV/c$, at the c.m. energy $\sqrt s=183\,\, GeV$.}
\label{Fig16} \vskip -0.2cm
\includegraphics[width=12.8cm]{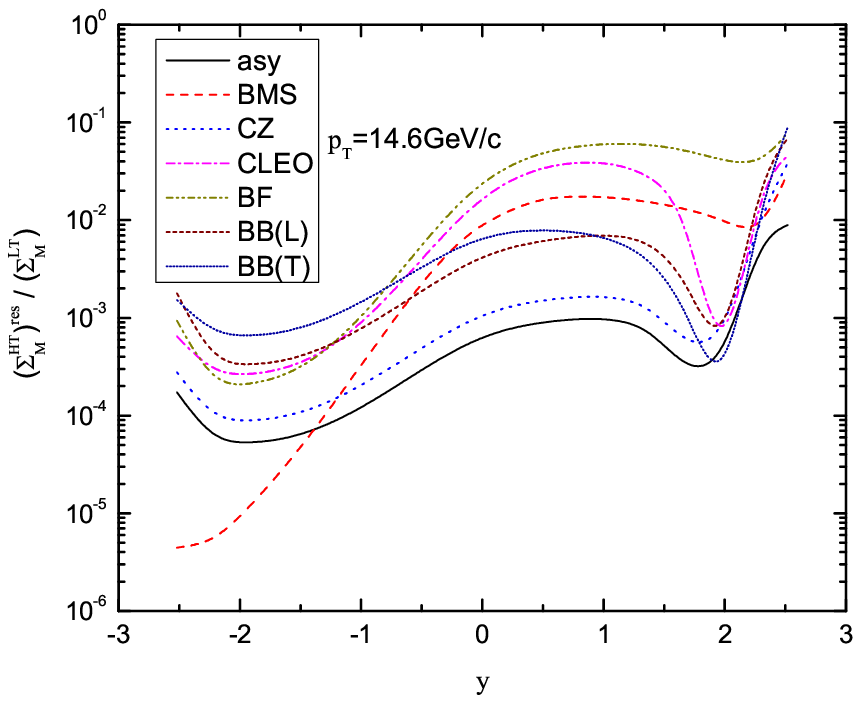}
\vskip-0.5cm \caption{Ratio
$(\Sigma_{M}^{HT})^{res}/(\Sigma_{M}^{LT})$, as a function of the
$y$ rapidity of the meson at the  transverse momentum of the meson
$p_T=14.6\,\, GeV/c$, at the c.m. energy $\sqrt s=183\,\, GeV$.}
\label{Fig17}
\end{figure}

\newpage

\begin{figure}[htb]
\includegraphics[width=12.8cm]{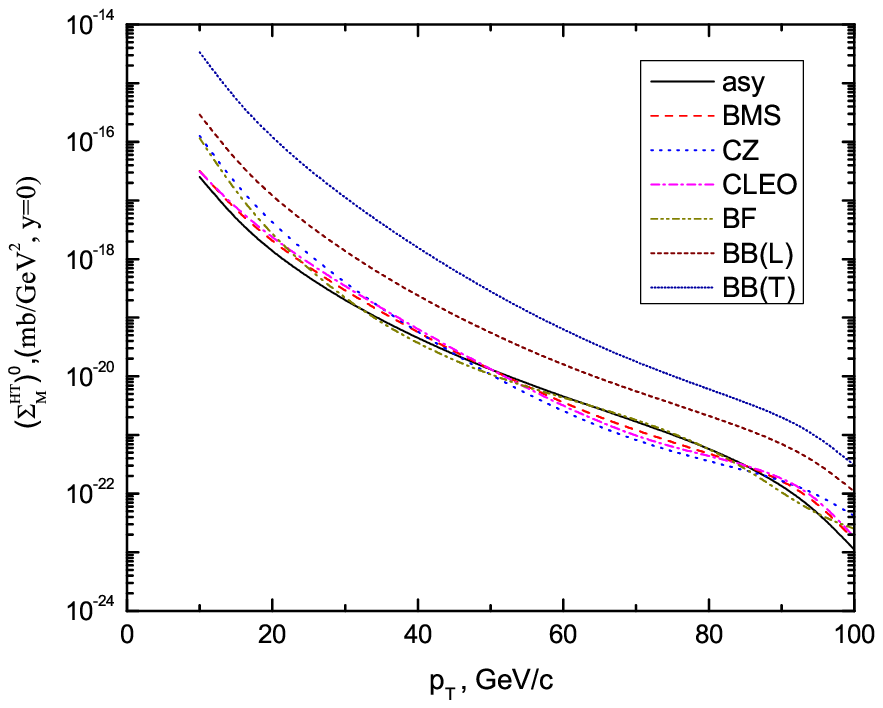}
\vskip-1.08cm \caption{Higher-twist $M$ production cross section
$(\Sigma_{M}^{HT})^{o}$ as a function of the $p_{T}$ transverse
momentum of the meson at the c.m.energy $\sqrt s=209\,\, GeV$.}
\label{Fig18} \vskip -0.2cm
\includegraphics[width=12.8cm]{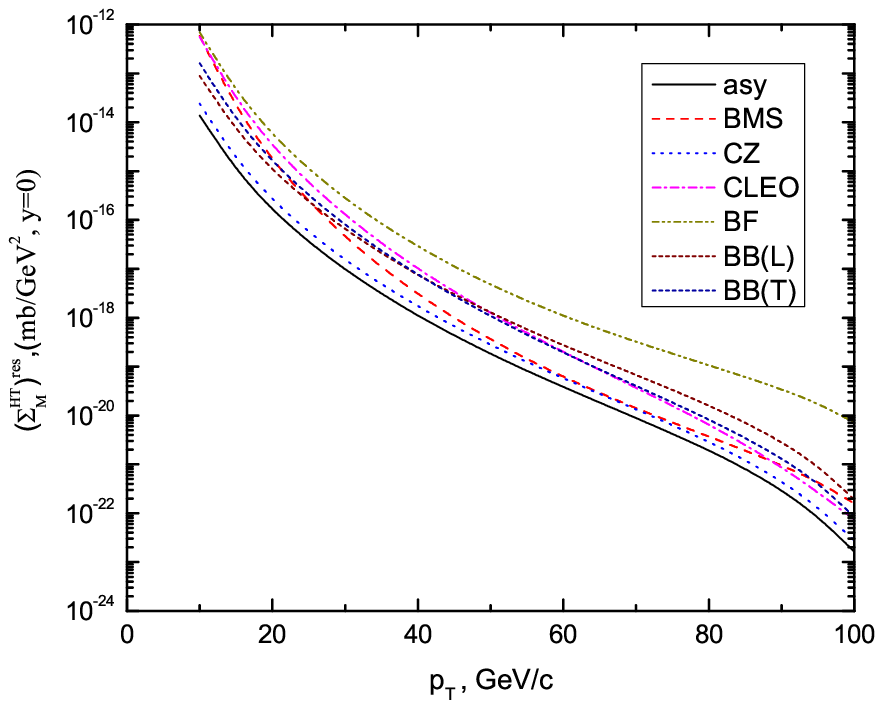}
\vskip-0.5cm \caption{Higher-twist $M$ production cross section
$(\Sigma_{M}^{HT})^{res}$ as a function of the $p_{T}$ transverse
momentum of the meson at the c.m.energy $\sqrt s=209\,\, GeV$.}
\label{Fig19}
\end{figure}

\newpage

\begin{figure}[htb]
\includegraphics[width=12.8cm]{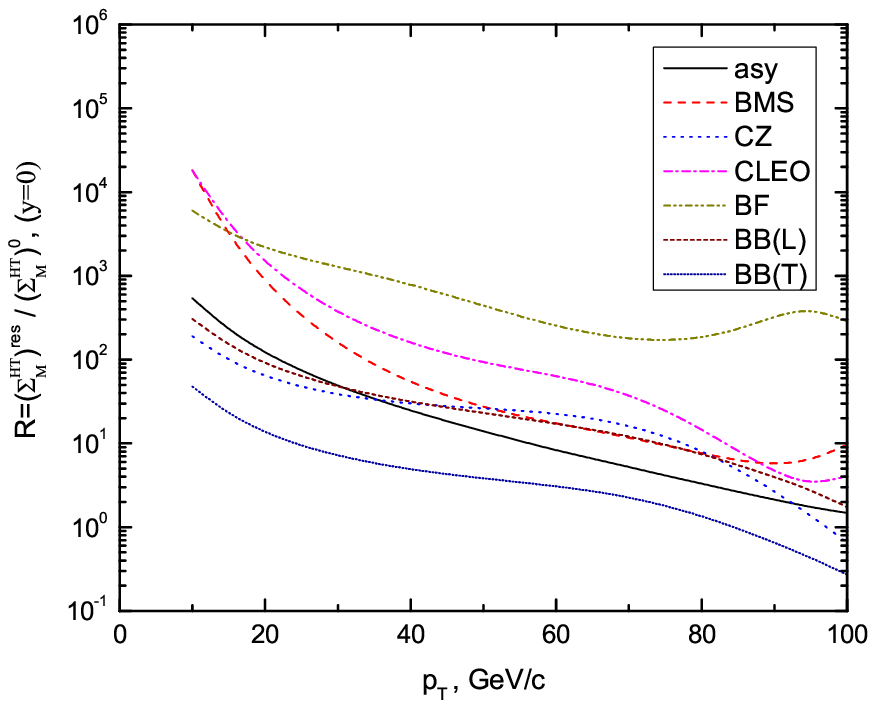}
\vskip-1.08cm \caption{Ratio
$R=(\Sigma_{M}^{HT})^{res}/(\Sigma_{M}^{HT})^{0}$, where
higher-twist contribution are calculated for the meson rapidity
$y=0$ at the c.m.energy $\sqrt s=209\,\,GeV$ as a function of the
meson transverse momentum, $p_{T}$.} \label{Fig20} \vskip -0.2cm
\includegraphics[width=12.8cm]{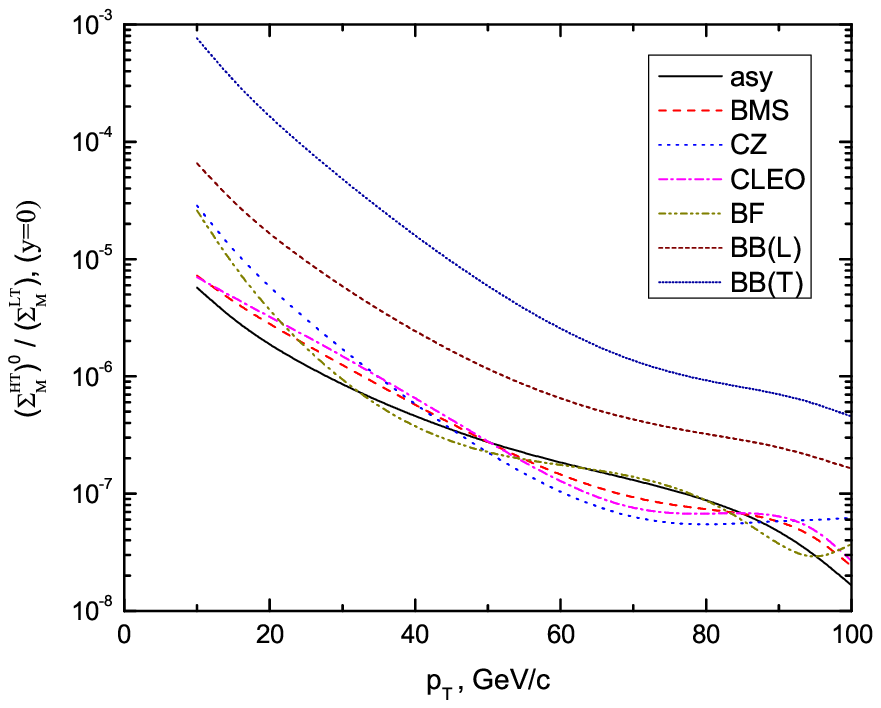}
\vskip-0.5cm \caption{Ratio
$(\Sigma_{M}^{HT})^{0}/(\Sigma_{M}^{LT})$, as a function of the
$p_{T}$ transverse momentum of the meson  at the c.m. energy
$\sqrt s=209\,\,GeV$.} \label{Fig21}
\end{figure}

\newpage

\begin{figure}[htb]
\includegraphics[width=12.8cm]{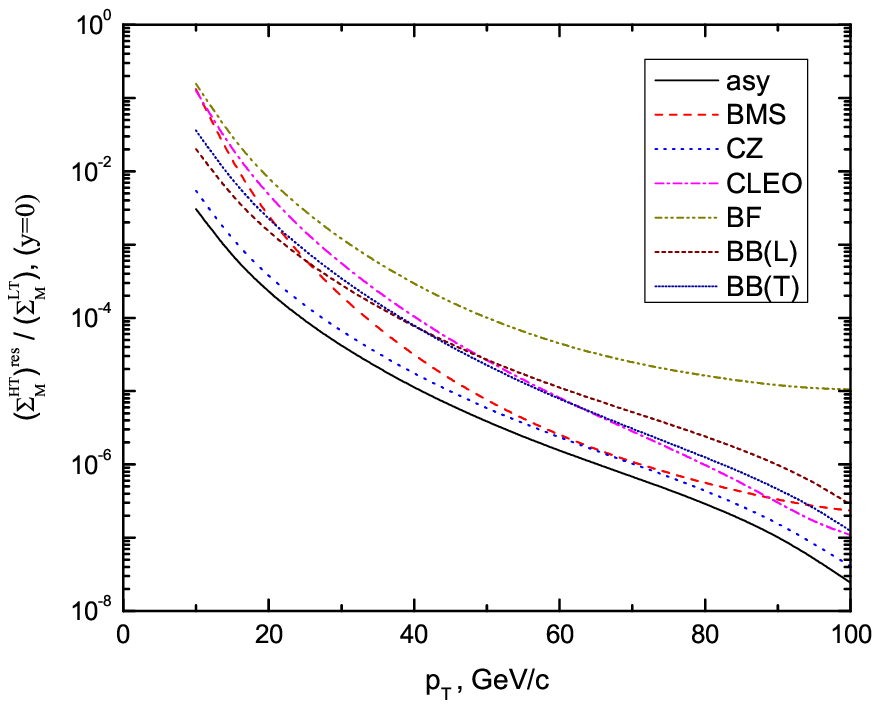}
\vskip-1.08cm \caption{Ratio
$(\Sigma_{M}^{HT})^{res}/(\Sigma_{M}^{LT})$, as a function of the
$p_{T}$ transverse momentum of the meson  at the c.m. energy
$\sqrt s=209\,\,GeV$..} \label{Fig22} \vskip -0.2cm
\includegraphics[width=12.8cm]{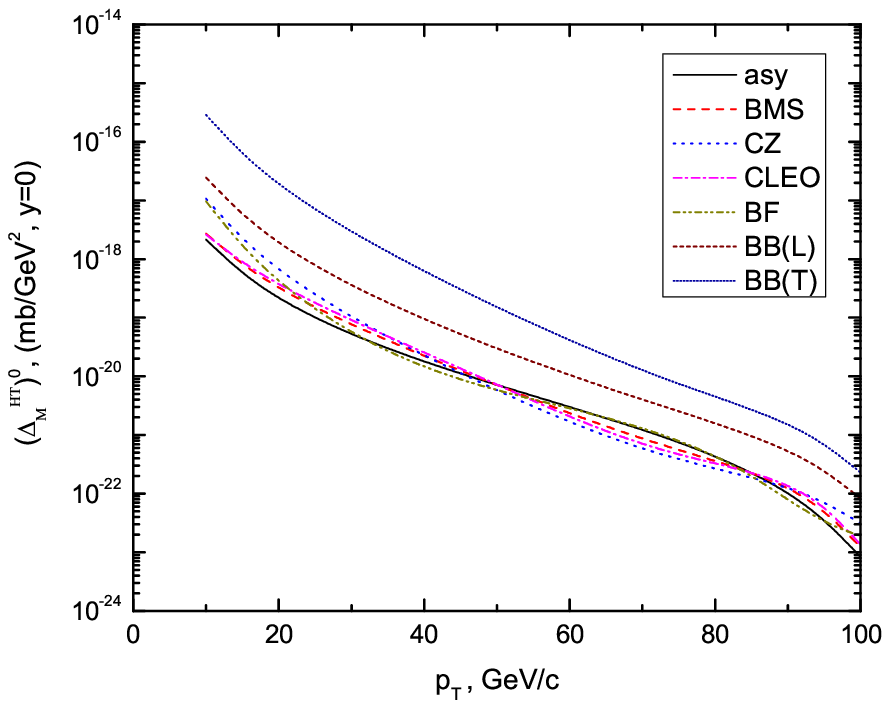}
\vskip-0.5cm \caption{The difference of the higher-twist cross
section,
$(\Delta_{M}^{HT})^{0}=(\Sigma_{M^{+}}^{HT})^{0}-(\Sigma_{M^{-}}^{HT})^{0}$,
as a function of the meson transverse momentum, $p_{T}$, at the
c.m.energy $\sqrt s=209\,\, GeV$.} \label{Fig23}
\end{figure}

\newpage

\begin{figure}[htb]
\includegraphics[width=12.8cm]{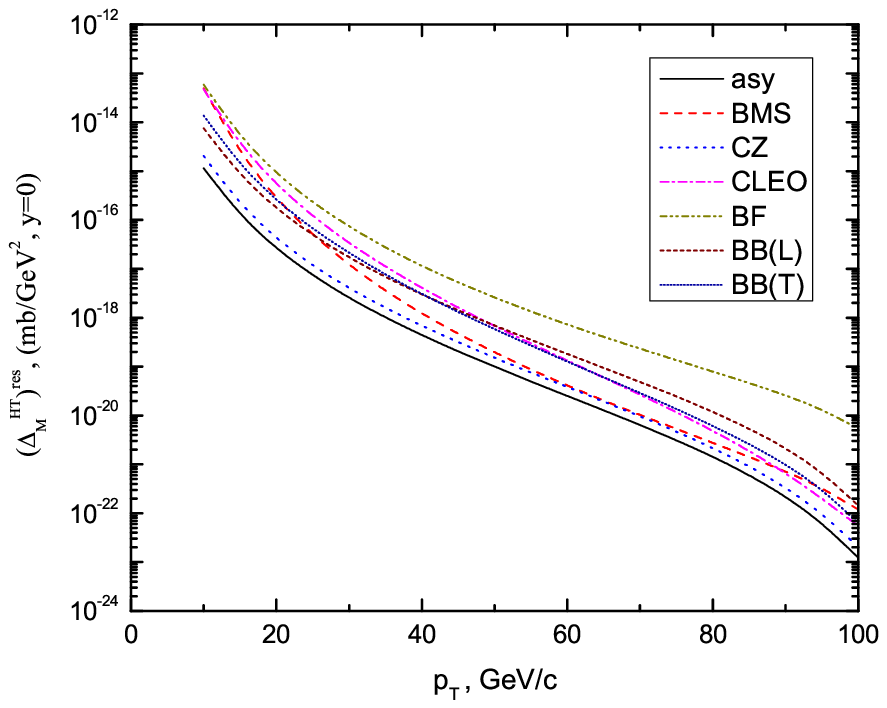}
\vskip-1.08cm \caption{The difference of the higher-twist cross
section,
$(\Delta_{M}^{HT})^{res}=(\Sigma_{M^{+}}^{HT})^{res}-(\Sigma_{M^{-}}^{HT})^{res}$,
\\ as a function of the meson transverse momentum, $p_{T}$, at the
c.m.energy $\sqrt s=209\,\, GeV$.} \label{Fig24} \vskip -0.2cm
\includegraphics[width=12.8cm]{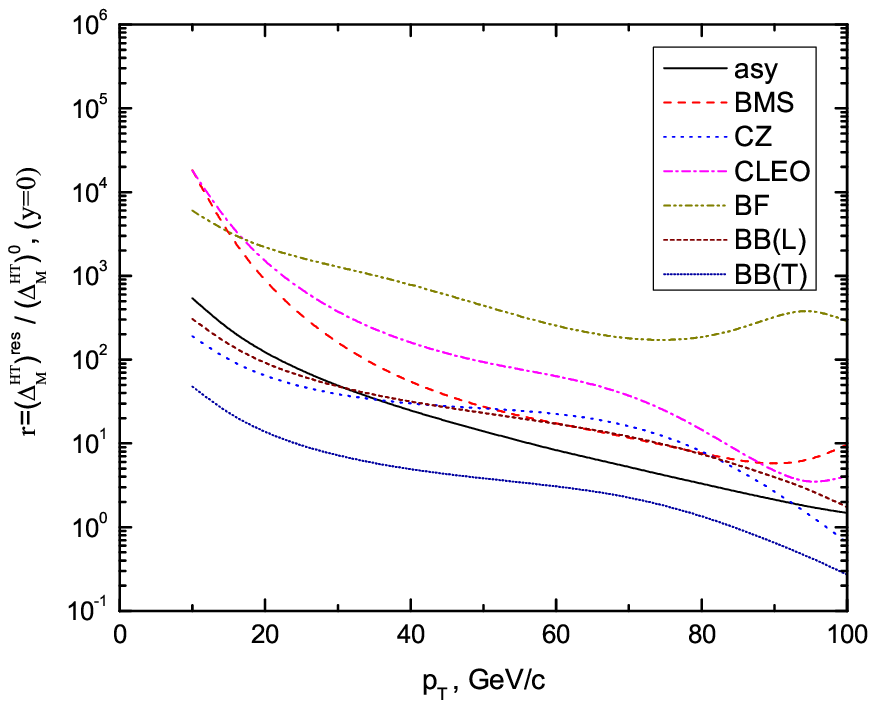}
\vskip-0.5cm \caption{Ratio
$r=(\Delta_{M}^{HT})^{res}/(\Delta_{M}^{HT})^{0}$, where higher-twist
contributions are calculated for the meson rapidity $y=0$ at
the c.m. energy $\sqrt s=209\,\, GeV$, as a function of the meson
transverse momentum, $p_T$.} \label{Fig25}
\end{figure}

\newpage

\begin{figure}[htb]
\includegraphics[width=12.8cm]{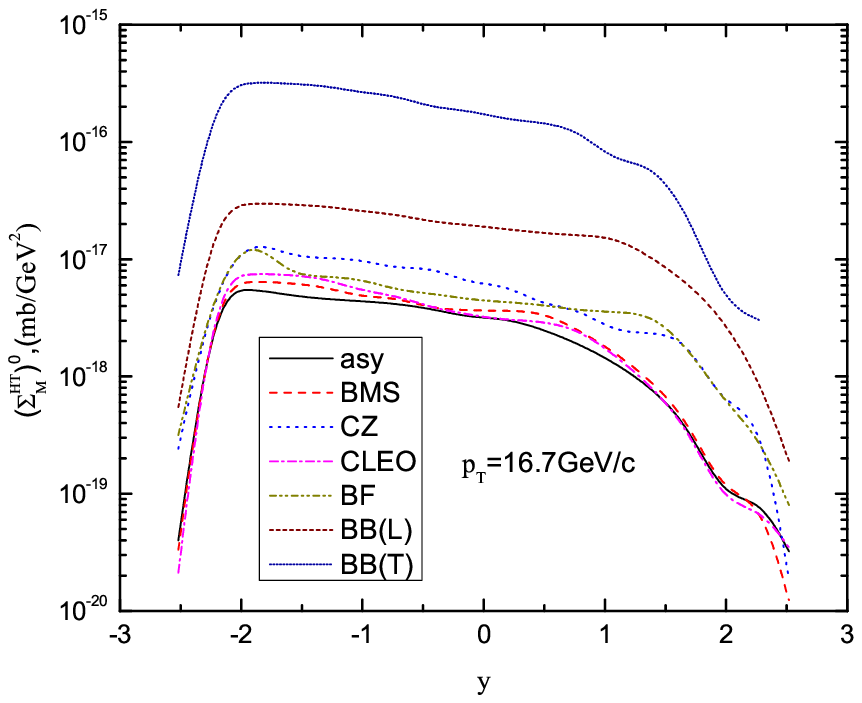}
\vskip-1.12cm \caption{Higher-twist $M$ production cross section
$(\Sigma_{M}^{HT})^{0}$, as a function of the $y$ rapidity of the
meson at the  transverse momentum of the meson $p_T=16.7\,\,
GeV/c$, at the c.m. energy $\sqrt s=209\,\, GeV$.} \label{Fig26}
\vskip-0.4cm
\includegraphics[width=12.8cm]{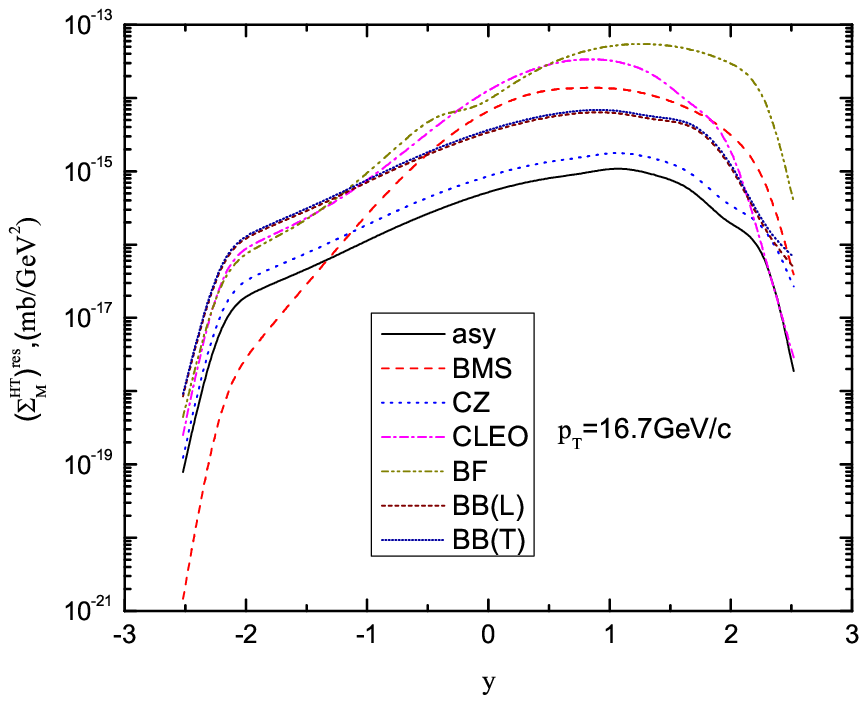}
\vskip-0.8cm \caption{Higher-twist $M$ production cross section
$(\Sigma_{M}^{HT})^{res}$, as a function of the $y$ rapidity of
the meson at the  transverse momentum of the meson $p_T=16.7\,\,
GeV/c$, at the c.m. energy $\sqrt s=209\,\, GeV$.} \label{Fig27}
\end{figure}

\newpage

\begin{figure}[htb]
\includegraphics[width=12.8cm]{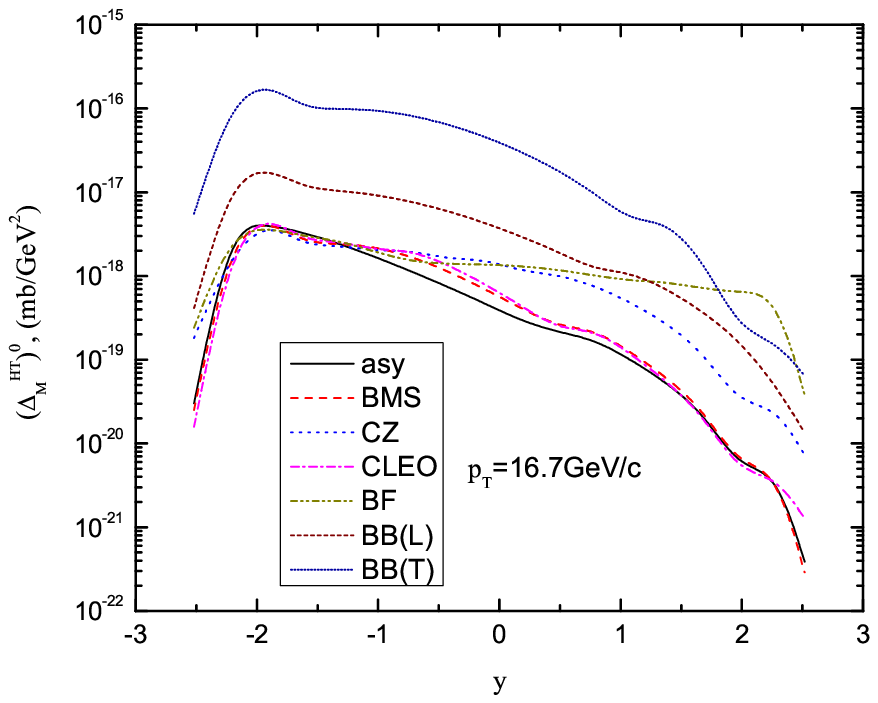}
\vskip-1.12cm \caption{The difference of the higher-twist cross
section,
$(\Delta_{M}^{HT})^{0}=(\Sigma_{M^{+}}^{HT})^{0}-(\Sigma_{M^{-}}^{HT})^{0}$,
as a function of the $y$ rapidity of the meson at the  transverse
momentum of the meson $p_T=16.7\,\, GeV/c$, at the c.m. energy
$\sqrt s=183\,\, GeV$.} \label{Fig28} \vskip -0.4cm
\includegraphics[width=12.8cm]{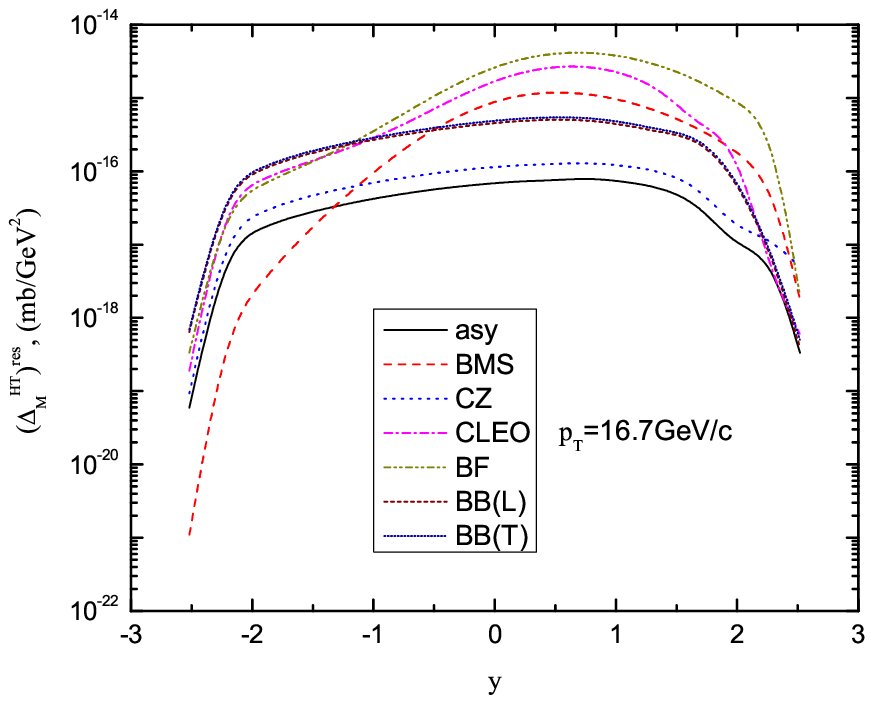}
\vskip-0.8cm \caption{The difference of the higher-twist cross
section,
$(\Delta_{M}^{HT})^{res}=(\Sigma_{M^{+}}^{HT})^{res}-(\Sigma_{M^{-}}^{HT})^{res}$,
as a function of the $y$ rapidity of the meson at the  transverse
momentum of the meson $p_T=16.7\,\, GeV/c$, at the c.m. energy
$\sqrt s=209\,\, GeV$.} \label{Fig29}
\end{figure}

\newpage

\begin{figure}[htb]
\includegraphics[width=12.8cm]{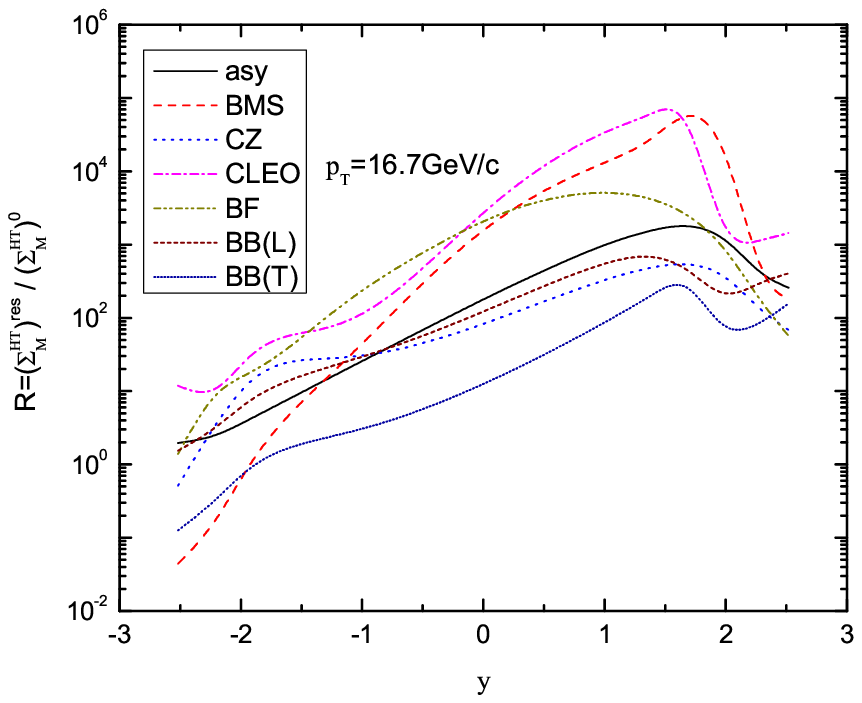}
\vskip-1.08cm \caption{Ratio
$R=(\Sigma_{M}^{HT})^{res}/(\Sigma_{M}^{HT})^{0}$, as a function
of the $y$ rapidity of the meson at the  transverse momentum of
the meson $p_T=16.7\,\, GeV/c$, at the c.m. energy $\sqrt
s=209\,\, GeV$.} \label{Fig30} \vskip -0.2cm
\includegraphics[width=12.8cm]{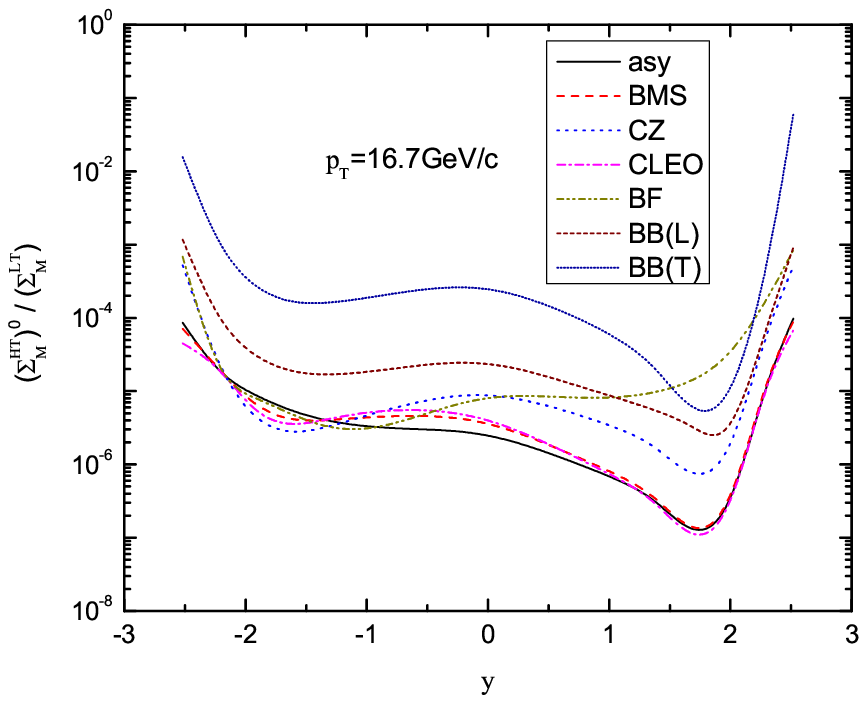}
\vskip-0.5cm \caption{Ratio
$(\Sigma_{M}^{HT})^{0}/(\Sigma_{M}^{LT})$, as a function of the
$y$ rapidity of the meson at the  transverse momentum of the meson
$p_T=16.7\,\, GeV/c$, at the c.m. energy $\sqrt s=209\,\, GeV$.}
\label{Fig31}
\end{figure}

\newpage

\begin{figure}[htb]
\includegraphics[width=12.8cm]{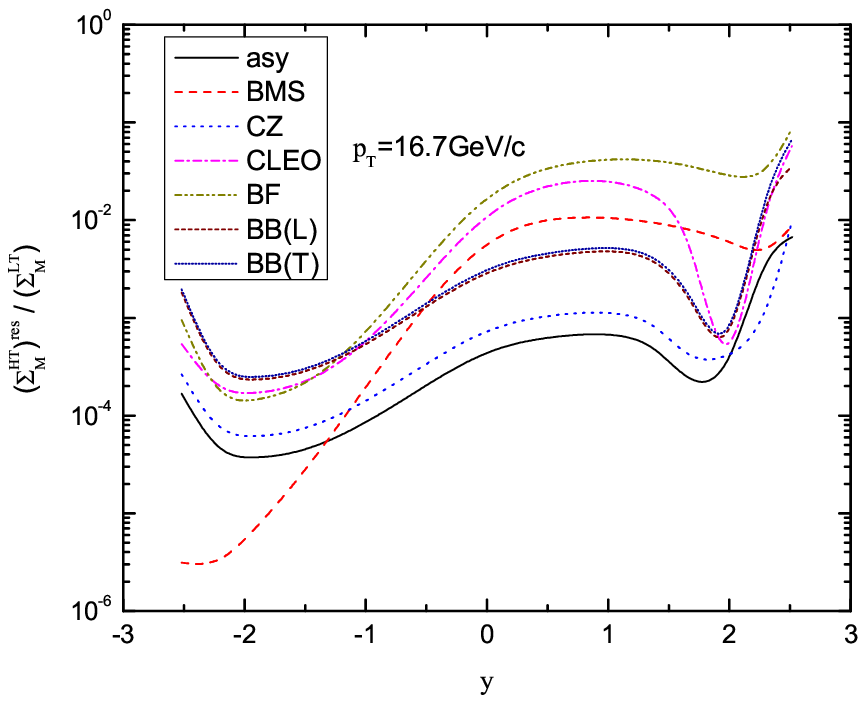}
\vskip-1.08cm \caption{Ratio
$(\Sigma_{M}^{HT})^{0}/(\Sigma_{M}^{LT})$, as a function of the
$y$ rapidity of the meson at the  transverse momentum of the meson
$p_T=16.7\,\,GeV/c$, at the c.m. energy $\sqrt s=209\,\, GeV$}.
\label{Fig32} \vskip -0.2cm
\includegraphics[width=12.8cm]{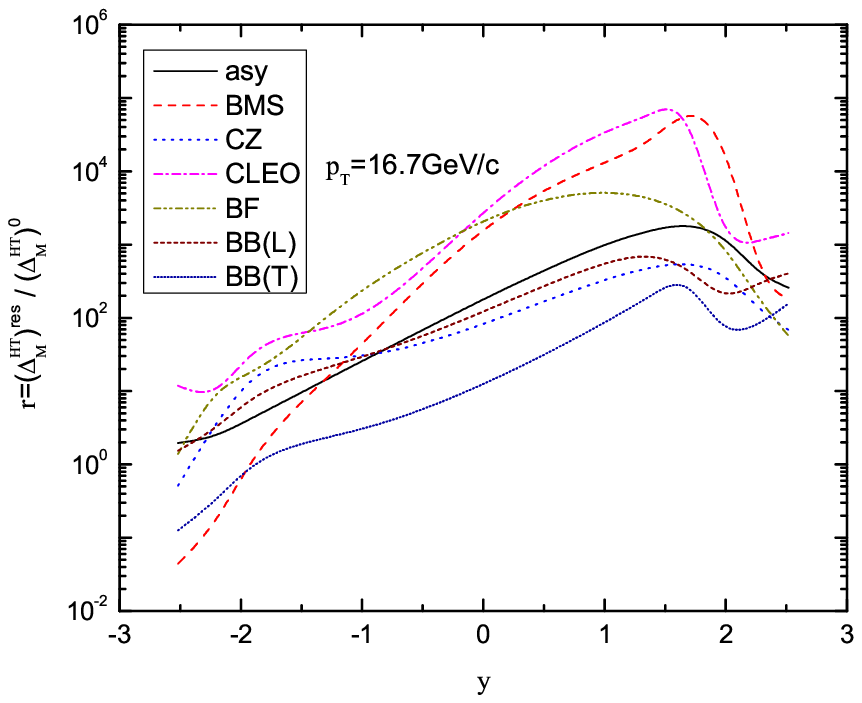}
\vskip-0.5cm \caption{Ratio
$r=(\Delta_{M}^{HT})^{res}/(\Delta_{M}^{HT})^{0}$, as a function
of the $y$ rapidity of the meson at the  transverse momentum of
the meson $p_T=16.7\,\, GeV/c$, at the c.m. energy $\sqrt
s=209\,\, GeV$} \label{Fig33}
\end{figure}

\begin{thebibliography}{99}
\section{References}
\bibitem{1}G. P.~Lepage and S. J. ~Brodsky, Phys. Lett. \textbf{87B}, 359 (1979); Phys. Rev. Lett. \textbf{43}, 545 (1979); \textbf{43}, 1625(E) (1979);
\bibitem{2}G. P.~Lepage and S. J.~Brodsky, Phys. Rev. \textbf{D22}, 2157 (1980).
\bibitem{3}A. V.~Efremov and A. V.~Radyushkin, Theor. Mat. Phys. \textbf{42}, 97 (1980); Phys. Lett. \textbf{94B}, 245 (1980).
\bibitem{4}A.~Duncan and A.~Mueller, Phys. Lett. \textbf{90B}, 159 (1980); Phys. Rev. \textbf{D21}, 1636 (1980).
\bibitem{5}A.V.~Radyushkin, Dubna Report No. P2-10717, 1977.
\bibitem{6} V. L.~Chernyak and A. R.~Zhitnitsky, Nucl. Phys. \textbf{B201}, 492 (1982).
\bibitem{7} V. L.~Chernyak and A. R.~Zhitnitsky, Nucl. Phys. \textbf{B246}, 52 (1984).
\bibitem{8} I. D.~King, C. T.~Sachrajda, Nucl. Phys. \textbf{B279}, 785 (1987).
\bibitem{9} V. L.~Chernyak, A. R.~Zhitnitsky, Phys. Rep. \textbf{112}, 173 (1984).
\bibitem{10} S. V.~Mikhailov and A. V.~Radyushkin, Pis'ma Zh. Eksp. Teor. Fiz. 43, 551 (1986)[JETP Lett. \textbf{43}, 712 (1986)]; Yad. Fiz. 49, 794 (1988)[Sov. J. Nucl. Phys. \textbf{49}, 494 (1989)]; Phys. Rev. \textbf{D45}, 1754 (1992).
\bibitem{11}V. M.~Braun and I. E.~Filyanov, Z. Phys. \textbf{C44}, 157 (1989).
\bibitem{12} G. R.~Farrar, K.~Huleihel and H.~Zhang, Nucl. Phys. \textbf{B349}, 655 (1991).
\bibitem{13}A. V.~Radyushkin and R.~Ruskov, Phys. Lett. \textbf{B374}, 173 (1996); Nucl. Phys. \textbf{B481}, 625 (1996).
\bibitem{14}S. J.~Brodsky and G. L.~Lepage, in Perturbative Quantum Chromodynamics, edited by A. H.~Mueller, (World Scientific, Singapore, 1989), p.93.
\bibitem{15}S. J.~Brodsky, H.-C.Pauli and S. S.~Pinsky, Phys. Rep. \textbf{301}, 299 (1998).
\bibitem{16}J.~Gronberg $et.al.$ (The CLEO Collaboration), Phys. Rev. \textbf{D57}, 33 (1998).
\bibitem{17} V. Yu.~Petrov, M. V.~Polyakov, R.~Ruskov, C.~Weiss and K.~Goeke, Phys. Rev. \textbf{D59}, 114018  (1999); hep-ph/9807229.
\bibitem{18}S. V.~Mikhailov, J.High Energy Phys. 06 (2007) 009.
\bibitem{19}V. A.~Matveev, R. M.~Muradyan, and A. N.~Tavkhelidze, Lett. Nuovo Cimento  \textbf{7}, 719 (1973).
\bibitem{20}S. J.~Brodsky and G.R.~Farrar, Phys. Rev. Lett. \textbf{31}, 1153  (1973).
\bibitem{21}J. F.~Gunion, S. J.~Brodsky, and R.~Blankenbacler,Phys. Rev. \textbf{D6}, 2652 (1972).
\bibitem{22}V. A.~Matveev, L. A.~Slepchenko, and A. N.~Tavkhelidze, Phys. Lett. \textbf{100B}, 75 (1981).
\bibitem{23}J. A.~Bagger and J. F.~Gunion, Phys. Rev. \textbf{D25}, 2287 (1982).
\bibitem{24}V. N.~Baier and A. G.~Grozin, Phys. Lett. \textbf{96B}, 181 (1980).
\bibitem{25}S. J.~Brodsky, G. P.~Lepage and P. B.~Mackenzie, Phys. Rev. \textbf{D28}, 228 (1983).
\bibitem{26}A. I.~Ahmadov, I.~Boztosun, A.~Soylu and E. A.~Dadashov, Int. J. Mod. Phys. \textbf{E17}, 1041 (2008); hep-ph/0611365.
\bibitem{27}M. Maul, E.~Stein, A.~Schafer, and L.~Mankiewich, Phys. Lett. \textbf{B401}, 100 (1997).
\bibitem{28}Y. L.~Dokshitzer, V. A.~Khoze and S. I.~Troyan, Phys. Rev. \textbf{D53}, 89 (1996).
\bibitem{29}A. L.~Kataev, Mod. Phys. Lett. A \textbf{20}, 2007 (2005).
\bibitem{30}G.'t Hooft, in The Whys of Subnuclear Physics, Erice, 1977, edited by A.~Zichnichi(Plenum, New York, 1979), p.94.
\bibitem{31}H.~Contopanagos and G.~Sterman, Nucl. Phys. \textbf{B419}, 77 (1994).
\bibitem{32}W.~Greiner, S.~Schramm, and E.~Stein, Quantum Chromodynamics (Springer,Berlin,2002), 2nd ed., p.551.
\bibitem{33}D. V.~Shirkov and I. L.~Solovtsov, Phys. Rev. Lett. \textbf{79}, 1209 (1997).
\bibitem{34}J. ~Botts and G.~ Sterman, Nucl.Phys. \textbf{B325}, 62 (1989).
\bibitem{35}I. V.~ Anikin, D. Yu.~ Ivanov, B.~Pire, L.~Szymanowski,
and S.~Wallon, Nucl. Phys. \textbf{B828}, 1 (2010); Phys. Lett.
\textbf{B682}, 413 (2010).
\bibitem{36}E.~Gardi, G.~Grunberg and M.~Karliner, J. High Energy Phys. 07 (1998) 007.
\bibitem{37}S. S.~Agaev, Phys. Lett. \textbf{B360}, 117 (1995); \textbf{B369}, 379(E) (1996);
Mod. Phys. Lett. \textbf{A10}, 2009 (1995);\textbf{11}, 957 (1996)
\bibitem{38}A. I.~Ahmadov, Coskun ~Aydin, Sh. M.~Nagiyev, Yilmaz A.~Hakan, and E. A.~Dadashov, Phys. Rev. \textbf{D80}, 016003 (2009).
\bibitem{39}V. M.~Budnev, I. F.~Ginzburg, G. V.~Meledin, and V. G.~Serbo, Phys.  Rep. \textbf{15}, 181 (1975).
\bibitem{40}J. A.~Bagger and J. F.~Gunion, Phys. Rev. \textbf{D29}, 40 (1984).
\bibitem{41}J. A.~Bagger and J. F.~Gunion, Phys. Rev. \textbf{D25}, 2287 (1982).
\bibitem{42}J. A.~Hassan and J. K.~Storrow, Z. Phys. \textbf{C14}, 65 (1982).
\bibitem{43}S. J.~Brodsky, T. A.~DeGrand, J. F.~Gunion and J. H.~Weis, Phys. Rev. Lett \textbf{41}, 672 (1978); Phys. Rev. \textbf{D19}, 1418 (1979).
\bibitem{44}G. P.~Lepage and S. J. ~Brodsky, Phys. Lett. \textbf{87B}, 359 (1979); Phys. Rev. Lett. \textbf{43}, 545 (1979); \textbf{43}, 1625(E) (1979); A.~Duncan and A.~Mueller, Phys. Rev. \textbf{D21}, 1636 (1980).
\bibitem{45}G. P.~Lepage and S. J.~Brodsky, Phys. Rev. \textbf{D22},  2157 (1980).
\bibitem{46}A.~Schmedding and O. Yakovlev, Phys. Rev. \textbf{D62}, 116002 (2000).
\bibitem{47} A. P.~Bakulev, S. V.~Mikhailov and N. G.~Stefanis, Phys. Lett, \textbf{B578}, 91 (2004).
\bibitem{48}S. V.~Mikhailov and A. V.~Radyushkin, Phys. Rev. \textbf{D45}, 1754 (1992).
\bibitem{49}A. P.~Bakulev, S. V.~Mikhailov, Z. Phys. \textbf{C68}, 451 (1995).
\bibitem{50}A. E.~Dorokhov, JETP Lett. \textbf{77}, 63 (2003).
\bibitem{51}P.~Ball and V. M.~Braun, Phys. Rev. \textbf{D54}, 2182 (1996); hep-ph/9602323.
\bibitem{52}S. S.~Agaev, Eur. Phys. J. C \textbf{1}, 321 (1998).
\bibitem{53}M.~Beneke and V. M.~Braun, Phys. Lett. \textbf{B348}, 513 (1995); P. Ball, M.~Beneke and V. M.~Braun, Nucl. Phys. \textbf{B452}, 563 (1995); M.~Beneke, Nucl. Phys. \textbf{B405}, 424 (1993).
\bibitem{54} J. Zinn-Justin, Phys. Rep. \textbf{70}, 109 (1981).
\bibitem{55}  A.~Erdelyi, Higher Transcendental Functions (McGrow-Hill, New York, 1953), Vol.2.
\bibitem{56} F. Cornet, Acta Phys. Polon. \textbf{B37}, 663 (2006); hep-ph/0601056.

\end{thebibliography}
\end{document}